\documentclass[referee,usenatbib]{mn2e}
\usepackage{rotating}
\usepackage{journals}
\usepackage{graphicx, subfigure}
\usepackage{xspace}
\usepackage{amssymb}
\usepackage{epsfig}
\usepackage{lscape}

\DeclareRobustCommand{\ion}[2]{%
\relax\ifmmode
\ifx\testbx\f@series
{\mathbf{#1\,\mathsc{#2}}}\else
{\mathrm{#1\,\mathsc{#2}}}\fi
\else\textup{#1\,{\mdseries\textsc{#2}}}%
\fi}

\def\Ha{H$\alpha$~}
\def\HI{\ion{H}{i}~} 
\def\HII{\ion{H}{ii}~} 
\def\HeII{\ion{He}{ii}~}
\def\CIV{\ion{C}{iv}~}  
\def\OI{\ion{O}{i}}
\def\OII{\ion{O}{ii}}
\def\NII{\ion{N}{ii}}
\def\SII{\ion{S}{ii}}
\def\deg{\hbox{$^\circ$}}
\def\kms{km~s$^{-1}$}
\def\ang{$\mathring{\mathrm{A}}$}
\title[\Ha imaging survey of Wolf-Rayet galaxies]{\Ha imaging survey of Wolf-Rayet galaxies: morphologies and star formation rates}
\author[Jaiswal \& Omar]{S. Jaiswal$^{1,2}$\thanks{E-mail: sumit@aries.res.in} 
and A. Omar$^{1}$\\ 
$^{1}$Aryabhatta Research Institute of Observational Sciences (ARIES), 
Manora Peak, Nainital, 263002, India\\
$^{2}$Pt. Ravishankar Shukla University, Raipur, 492010, India}

\begin{document}
\date{Accepted ---. Received ---; in original form ---}

\pagerange{\pageref{firstpage}--\pageref{lastpage}} \pubyear{2014}

\maketitle

\label{firstpage}
\begin{abstract}

The \Ha and optical broadband images of 25 nearby Wolf-Rayet (WR) galaxies are presented. The WR galaxies are known to have the presence of a recent ($\le$10 Myr) and massive star formation episode. The photometric \Ha fluxes are estimated, and corrected for extinction and line contamination in the filter pass-bands. The star formation rates (SFRs) are estimated using \Ha images and from the archival data in the far-ultraviolet (FUV), far-infrared (FIR) and 1.4 GHz radio continuum wave-bands. A comparison of SFRs estimated from different wavebands is made after including similar data available in literature for other WR galaxies. The \Ha based SFRs are found to be tightly correlated with SFRs estimated from the FUV data. The correlations also exist with SFRs estimates based on the radio and FIR data. The WR galaxies also follow the radio-FIR correlation known for normal star forming galaxies, although it is seen here that majority of dwarf WR galaxies have radio deficiency. An analysis using ratio of non-thermal to thermal radio continuum and ratio of FUV to \Ha SFR indicates that WR galaxies have lesser non-thermal radio emission compared to normal galaxies, most likely due to lack of supernova from the very young star formation episode in the WR galaxies.  The morphologies of 16 galaxies in our sample are highly suggestive of an ongoing tidal interaction or a past merger in these galaxies. This survey strengthens the conclusions obtained from previous similar studies indicating the importance of tidal interactions in triggering star-formation in WR galaxies.
\end{abstract}

\begin{keywords}
galaxies: interaction -- galaxies: starburst -- galaxies: Wolf-Rayet -- galaxies: star formation rates -- galaxies: \Ha photometry
\end{keywords}

\section{Introduction}

Wolf-Rayet (WR) galaxies are a subset of emission-line and \HII galaxies whose integrated optical spectra exhibit broad emission line features (\HeII $\lambda$4686, \CIV $\lambda$5808, etc.: Allen et al. 1976; Osterbrock \& Cohen 1982; Conti 1991), originating in the stellar winds of WR stars. The line strengths of the WR features indicate a substantial population ($10^2$ - $10^5$) of WR stars in these galaxies (e.g., Kunth \& Sargent 1981; Kunth \& Schild 1986). The most massive O-type stars (M $\ge$ 25 M$_\odot$ for solar metallicity) become WR stars after 2 to 5 Myr from their birth, spending only a short time (t$_{WR} \le$ 0.5 Myr) in this phase (Meynet \& Maeder 2005), before ending through supernova explosions. The WR stars are short-lived massive stars, therefore, their presence in a large number is a strong indicator of a young ($<$10 Myr) ongoing star-burst in a galaxy (e.g., Schaerer et al.\ 1999). The WR galaxies, therefore, offer an unique opportunity to study the onset of star formation as well as the conditions, which may be responsible for triggering star-formation in galaxies (Schaerer \& Vacca 1998). WR galaxies are quite rare in the nearby Universe as the WR phase is a very short-lived stage. The first identification of the WR features in a galaxy was in the blue compact dwarf galaxy He 2-10 (Allen et al. 1976). Conti (1991) compiled a WR galaxy catalogue having 37 objects. Later, Schaerer et al. (1999) made a catalogue of 139 WR galaxies. The most recent catalogue of WR galaxies is compiled by Brinchmann et al. (2008) using the $Sloan~Digital~Sky~Survey ~(SDSS)$ data. It lists 570 WR galaxies and a further 1115 suspected WR galaxies. Gil de Paz et al. (2003) presented $B$, $R$ and \Ha images of a total of 114 nearby blue compact dwarf (BCD) galaxies, many of those were identified as WR galaxies with high star formation rates.

The star-formation is considered a fundamental parameter for evolution of galaxies. The star-formation in a galaxy is measured in terms of star formation rate (SFR). The SFR is estimated from the number of Lyman continuum photons ($N_{LyC}$), which are emitted mainly by the young massive stars. The $N_{LyC}$ can be measured from the continuum flux in far ultraviolet (FUV) or blue wave-bands (e.g., Gallagher et al. 1984; Bell \& Kennicutt 2001; Salim et al. 2007; Murphy et al. 2011), far-infrared (FIR, e.g., Kennicutt et al. 1987; Misiriotis et al. 2004; Calzetti et al. 2010) wave-bands or radio wave-bands (e.g., Condon 1992; Cram et al. 1998; Garn et al. 2009; Murphy et al. 2011). The optical line emissions such as \Ha $\lambda$6563 and [\OII] $\lambda$3727 (e.g., Kennicutt \& Kent 1983; Gallego et al. 1995; Young et al. 1996; Moustakas et al. 2006; Prabhu 2012) from ionized gas also provide an estimate for SFR in galaxies. The \Ha emission directly provides an estimate of the ionizing flux from the young massive stars, and is amongst the strongest optical emission lines in star forming galaxies. The \Ha emission flux is less sensitive to dust extinction and metallicity compared to the [\OII] line. Other indicators such as X-ray continuum (e.g., Ranalli et al. 2003; Mineo et al. 2011) and mid-infrared PAH (polycyclic aromatic hydrocarbon) emission (e.g., Roussel 2001; Calzetti et al. 2007; Calzetti 2011) can also be used to get an estimate for SFR in galaxies. 

These SFR indicators often predict different values of SFR due to various uncertainties arising from extinction correction, stellar evolution models, magnetic field strengths, diffusion and re-processing time scales of stellar continuum photons in galaxies (Calzetti 2013). The FUV continuum emission from the young and massive stars can be sensitive to star formation events over the time scales of a few tens of Myr. The diffuse synchrotron radio emission tracing its origin in loss of energy from relativistic cosmic electrons accelerated in supernovae events and diffused out to kilo-parsec scales is sensitive to star formation events typically during the last 100 Myr. The \Ha emission line, on the other hand, provides information on the most recent ($\sim$10~Myr) star formation in galaxies (Kennicutt 1998a; Murphy et al. 2011). In case of  star formation taking place in highly dust-obscured regions, accuracy of extinction correction becomes very important in getting SFR estimates from the FUV and optical bands.  Although, the radio emission is not affected by dust, uncertainties on strength of magnetic field and contamination due to active galactic nuclei (AGN) may make radio flux a relatively less reliable estimator for the SFR in galaxies. Investigations on correlations between different SFR tracers are important to understand various physical processes in the interstellar medium (ISM). Such correlations also help in constraining SFRs in galaxies. Due to very recent star formation episode in WR galaxies, the star formation age may have subtle effects on the observable fluxes at different wavebands, and hence on SFR correlations. Such effects have not been studied very well over a large range of SFR values in WR galaxies. In a study of 20 WR galaxies, L\'opez-S\'anchez (2010) found that SFR derived from the \Ha photometry of WR galaxies yield statistically significant correlations with other tracers such as FUV, FIR, X-ray and radio continuum. 
 
The star formation trigger in WR galaxies is also important to understand. Several studies on normal star-forming galaxies indicate that gravitational tidal interactions and mergers of galaxies play a major role in galaxy evolution and triggering star formation in galaxies (Larson \& Tinsely 1978; Koribalski 1996; Kennicutt 1998b; Nikolic et al. 2004; Joseph \& Wright 1985; Solomon \& Sage 1988; Sanders \& Mirabel 1996; Genzel et al. 1998; Omar \& Dwarakanath 2005). The hierarchical growth models of galaxies support the formation of larger structures like giant spiral and massive elliptical galaxies through mergers and accretion of smaller structures like dwarf galaxies (e.g., Shlosman 2013; Amorisco et al.\ 2014; Deason et al.\ 2014), implying that tidal interactions should be fairly common. A causal link between star formation and mergers or interactions has been seen in several blue compact dwarf galaxies (e.g., Bravo-Alfaro et al.\ 2004; Bekki 2008; L{\'o}pez-S{\'a}nchez et al. 2012; Ashley 2014). The tidal interactions between galaxies can be traced in several ways. The morphological features such as tidal tails and plumes detected in the \HI 21cm-line images or in the optical images are a clear sign of interaction. Such tidal features are more easily detectable in the \HI 21cm-line compared to the optical images (L\'opez-S\'anchez 2010; L\'opez-S\'anchez et al. 2012; Lelli et al. 2014; McQuinn et al. 2015). The mergers can be detected in the optical bands through identification of multiple nuclei and arcs like features. Significant differences in the chemical composition of different star-forming regions within the same galaxy is also a sign of interaction (e.g., L\'opez-S\'anchez et al. 2004a,2004b,2006; L{\'o}pez-S{\'a}nchez \& Esteban 2009; L\'opez-S\'anchez 2010). Due to interaction induced star formation, the nuclei or arcs can be easily detected in the \Ha line imaging. A misalignment between the \Ha disk and the stellar disk can also be an indicator of interaction as seen in the WR dwarf galaxy MRK~996 (Jaiswal \& Omar 2013). The interaction features in galaxies can also be traced by studying kinematics of the ionized gas or neutral \HI gas (e.g., L\'opez-S\'anchez et al. 2004a,2004b,2006; L{\'o}pez-S{\'a}nchez \& Esteban 2009; L\'opez-S\'anchez 2010). 

Despite the importance of studying interactions in WR galaxies, very few detailed studies of WR galaxies have been carried out. M\'endez \& Esteban (2000) performed deep optical imaging and spectroscopy on a sample of WR galaxies, and found that the star-bursts triggered in low mass WR galaxies could be due to interactions with dwarf galaxies. L{\'o}pez-S{\'a}nchez \& Esteban (2008, 2009, 2010a, 2010b) and L{\'o}pez-S{\'a}nchez (2010) found that majority of galaxies in a sample of 20 WR galaxies were clearly interacting or merging with low luminosity dwarfs objects or intergalactic \HI clouds. It is suggested by these authors that interacting or merging nature of WR galaxies can be detected only when both deep, high-resolution images and spectroscopy data are available, because the interaction features are often faint. Karthick et al. (2014) recently studied properties of star-forming regions in a sample of seven WR galaxies through their broad-band and narrow-band photometry along with low-resolution optical spectroscopy. They constrained the age of the most recent star-formation events as 3-6 Myr. Four galaxies in their sample also show clear interaction features.  There are only a few other studies where tidal interactions were inferred in individual WR galaxies (e.g., Esteban \& M\'endez 1999; M\'endez \& Esteban 1999; M\'endez et al. 1999a, 1999b; Jaiswal \& Omar 2013). There is clearly a need to expand these studies to a larger number of WR galaxies.

The \Ha and $r$-band images of 25 WR galaxies are presented here. A total of 23 out of 25 WR galaxies in the present sample were selected from SDSS, which provides broad-band photometry and spectroscopic data in the visible range of the electromagnetic spectrum (see, Abazajian et al. 2009 and the references therein). Since the SDSS spectrum has been obtained using a fiber slit of $3''$ diameter and galaxy sizes of WR galaxies are several arc-min in our sample, total SFR can not be reliably derived from the SDSS spectroscopic data alone. The narrow-band \Ha photometry maps full extent of galaxies, and therefore, provides total SFR. The \Ha flux calculated from the photometric observations needs to be corrected for contaminations by other emission lines in the filter passbands apart from the galactic and internal extinction corrections. These corrections need a knowledge of the physical conditions in the ISM, and hence require spectroscopic data. The corrections to the \Ha flux were estimated in this paper primarily using the SDSS spectroscopic data. We have combined our sample of 25 WR galaxies with the database on 20 WR galaxies from L\'opez-S\'anchez (2010) making the total sample size up to 45 galaxies. The H$\alpha$-derived SFR estimates are compared with those estimated from the FUV, FIR and radio continuum luminosities from the archival data. The radio-FIR correlation is also constructed using the archival data from {\it IRAS (Infrared Astronomical Satellite)} at FIR wavelengths, and the {\it FIRST (Faint Images of the Radio Sky at Twenty-cm)} and the {\it NVSS (NRAO VLA Sky Survey)} images at 1.4~GHz taken with the Very Large Array (VLA). The \Ha and SDSS images were used to investigate interaction features in WR galaxies.

\section{Sample Properties}

The WR features are found in nearly all morphological types of galaxies, ranging from low-mass dwarf galaxies and irregular galaxies to massive spirals, luminous mergers, infrared luminous galaxies and Seyfert galaxies (Ho et al. 1995; Heckman \& Leitherer 1997; Zhang et al. 2007). We used the catalogs of WR galaxies prepared by Schaerer et al. (1999) and Brinchmann et al. (2008) to construct a sample. We selected galaxies up to $\sim$25 Mpc distance (with $H_0 = 75~\rm km~s^{-1} Mpc^{-1}$), so that galaxies can be observed with the \Ha filters available on the telescopes used. We further restricted our sample with the declination $> - 25\deg$, so that galaxies can be observed for sufficient integration time from the telescopes. We finally selected a total of 25 galaxies to carry out this study. The selected galaxies are brighter than 17 mag in the SDSS $r$-band, except NGC~2799 and UM~3111 whose SDSS $r$ magnitudes are 18.6 and 17.8 respectively. 

\begin{figure*}
\centering
\includegraphics[angle=0,width=0.48\linewidth]{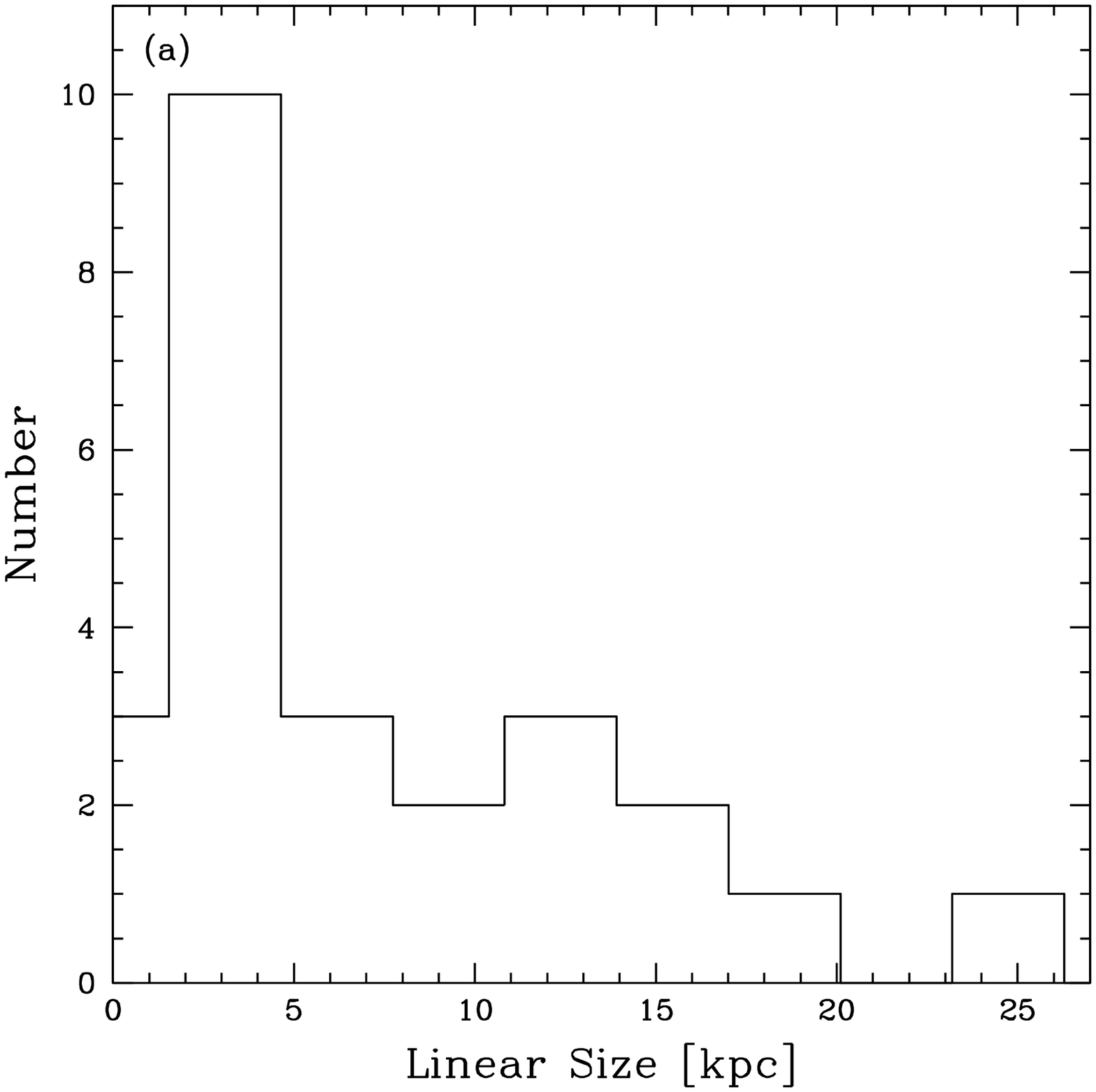}
\includegraphics[angle=0,width=0.48\linewidth]{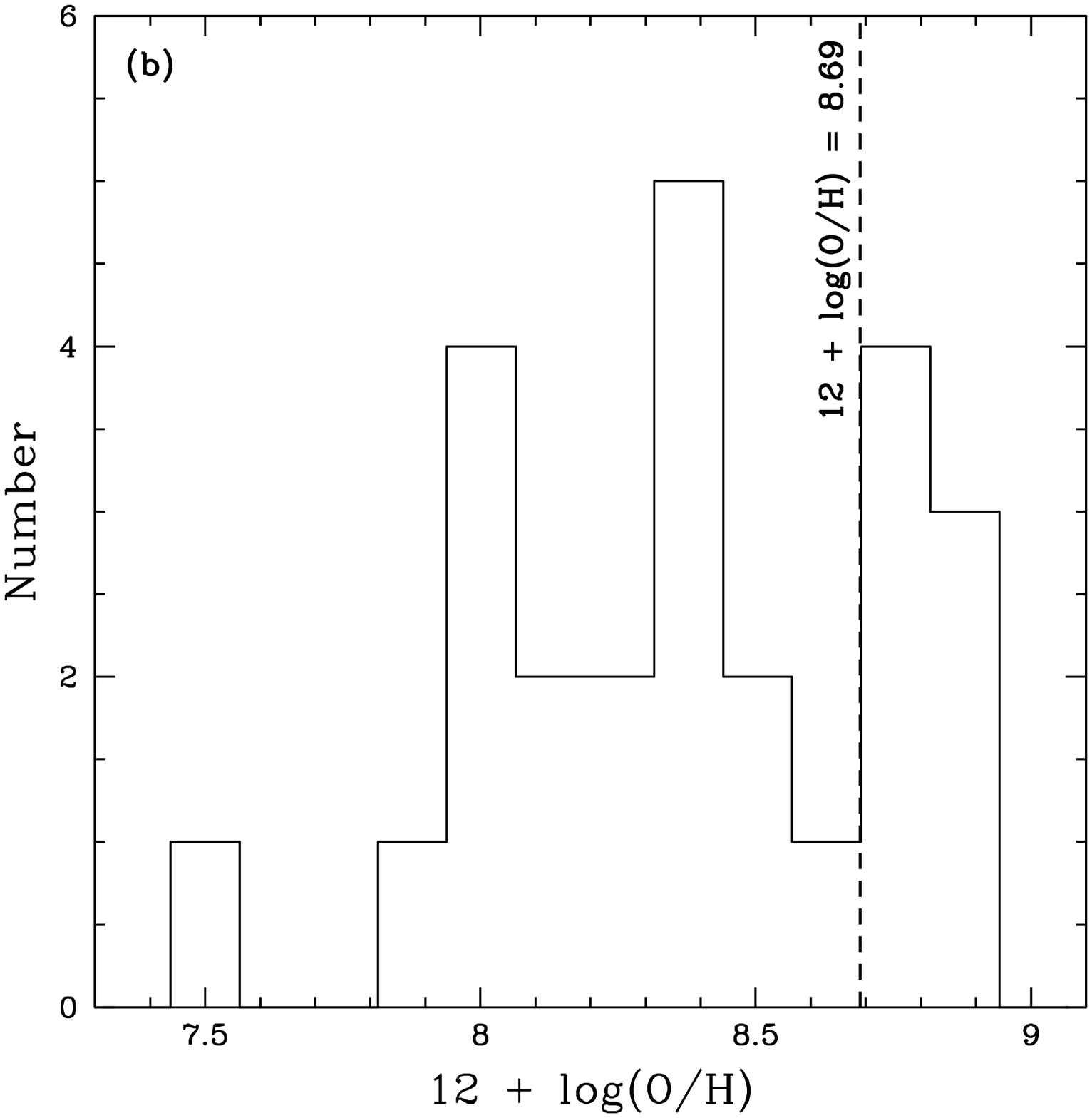}
\caption[ ]{The histograms for size and metallicity of galaxies in our sample of WR galaxies. The vertical dashed line in the metallicity plot shows the solar metallicity, 12 $+$ log(O/H) $=$ 8.69.}
\label{histogram}
\end{figure*}

\begin{figure*}
\centering
\includegraphics[angle=0,height=23cm,width=17cm]{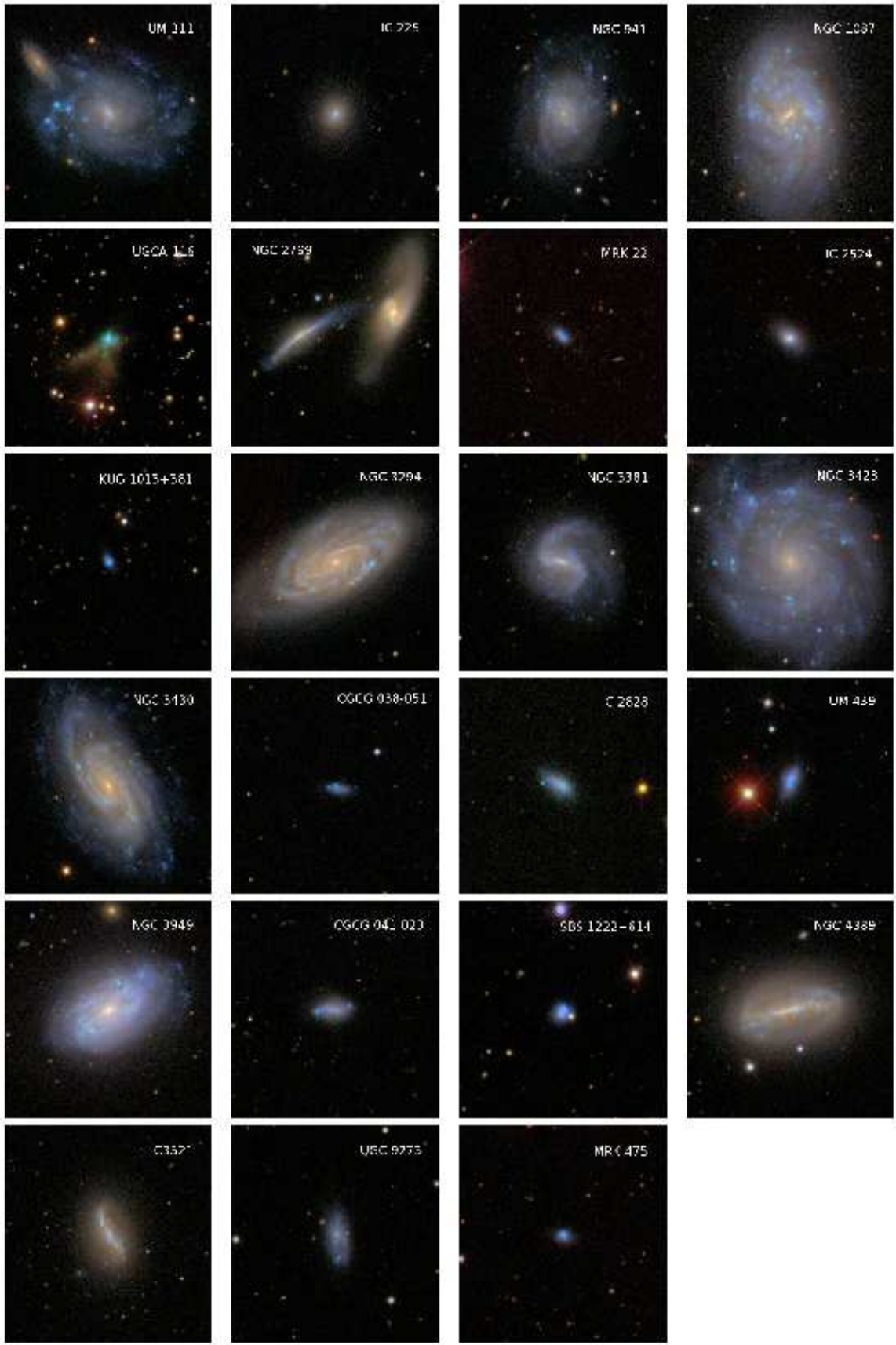}
\caption{SDSS color images of galaxies in our sample. These images are made using $g$, $r$ and $i$ bands. Each image has a size of 3.4$'$. North is up and East is to the left.}
\label{sdss}
\end{figure*}

The basic properties of the galaxies in our sample are given in Table~\ref{sample}, which is constructed using NASA/IPAC extragalactic database (NED). In this table, the SDSS $r$-band magnitudes for galaxies UGCA~116, UGCA~130, UM~439 and I~SZ~59 are extrapolated from the Johnson $B$ and Cousins $R$ band magnitudes (Gil de Paz et al. 2003) with the help of Lupton transformation relations published on the SDSS DR4 website, as these galaxies were not observed in the SDSS. The extinction and metallicity parameters for the selected galaxies are given in Table~\ref{archive}. For most of the galaxies, the mixed oxygen abundance estimates as given in Brinchmann et al. (2008) are provided here. Figure~\ref{histogram} shows the histograms of linear sizes and oxygen abundances of the selected galaxies. The linear size distribution indicates that the sizes of galaxies in the sample are $<$25 kpc, and majority of them are $<$10 kpc. We term galaxies with size $<$10 kpc as small and other galaxies as large. The oxygen abundance histogram shows that the selected WR galaxies have metallicities between 8 and 9. A vertical line showing the solar metallicity, 12 $+$ log(O/H) $=$ 8.69 (Asplund et al. 2009) is drawn in this figure. The majority of galaxies in our sample have sub-solar metallicities. The color composite images made using the SDSS $g$, $r$ and $i$ bands images of the galaxies are shown in Figure~\ref{sdss}. The blue regions here show star forming regions.

\section{Observations}

The \Ha and $r$-band observations of 21 galaxies were carried out using the 1.3-meter Devasthal Fast Optical Telescope (DFOT; longitude $= 79^{o}41'04''$~E, latitude $= 29^{o}21'40''$~N, altitude $\sim$2420~m above the mean sea level) near Nainital, India. This telescope has the Ritchey-Chr\'etien (RC) optical system with f/4 Cassegrain focus providing a plate scale of $0.54''$ pixel$^{-1}$. The data were recorded on a thermo-electrically cooled ($-80$\deg C) CCD (charge coupled device) camera having a back-illuminated E2V chip of $2048\times 2048$ pixels of size 13.5 $\mu$m. The CCD covers a circular field of view of diameter $\sim 18'$ on the sky. The CCD readout noise is nearly 7~$e^{-}$ at 1 MHz speed with a gain of 2~$e^{-}ADU^{-1}$. A brief overview of this telescope is provided in Sagar et al. (2011) and Jaiswal \& Omar (2013). The observations of the remaining 4 galaxies were carried out using the 2-meter telescope of the Inter University Centre for Astronomy and Astrophysics (IUCAA) Girawali Observatory (IGO; longitude $= 73^{o}40'$~E, latitude $= 19^{o}5'$~N, altitude $\sim$1000~m; see Das et al. 1999), Pune, India. This telescope also has a RC optical design with f/10 Cassegrain focus. The data were recorded using IUCAA Faint Object Spectrograph and Camera (IFOSC), which uses an E2V 2K$\times$2K, thinned, back-illuminated CCD with 13.5$\mu$m pixel size. The CCD used for imaging provides a field of view of $10.5'\times 10.5'$ on the sky corresponding to a plate scale of $0.3''$ pixel$^{-1}$. The gain and readout noise of the CCD camera are 1.5 $e^{-}ADU^{-1}$ and 4$e^{-}$, respectively. 

\begin{figure*}
\centering
\includegraphics[angle=-90,width=0.8\linewidth]{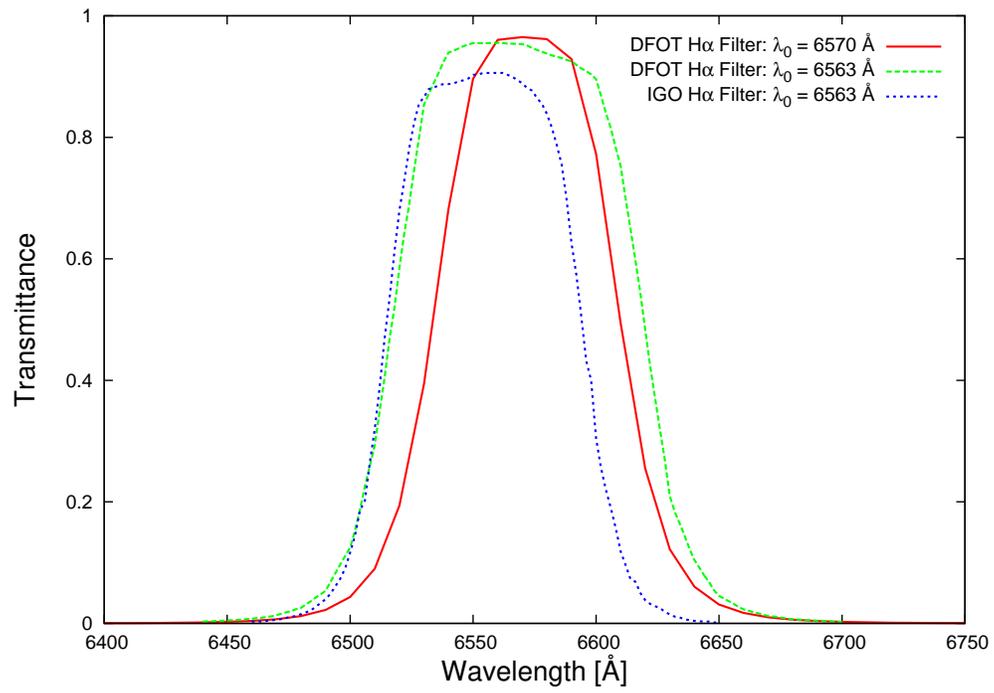}
\caption[ ]{Transmittance vs. wavelength plot for different \Ha filters used in the present study.}
\label{filters}
\end{figure*}

The observations were carried out mostly in the dark nights near the new moon periods. The observation log is provided in Table~\ref{obs}. The observations were carried out in the \Ha narrow-band filter and the SDSS-$r$ or Cuisine $R$ filter available on the telescopes. The central wavelength and FWHM (Full Width Half Maxima) of the \Ha filters used, and \Ha sensitivities in the observations are given in Table~\ref{cal}. Here, the \Ha sensitivity was calculated for 3$\sigma$ detection, and $\lambda_{\mathrm{H}\alpha}$ is the observed wavelength for the red-shifted H$\alpha$ emission and $\lambda_0$ is the central wavelength of the \Ha filter. The pass-band transmittance curves for the \Ha filters used are shown in Figure~\ref{filters}. The total integration time was divided into $\sim$25 frames of durations between 5 and 10 minutes. The total integration time in the \Ha band was typically between 100 and 200 minutes while that in the $r$ band was typically between 10 and 30 minutes. At least one standard spectrophotometric star selected from Oke (1990) was observed a few times at different airmass values every night. The standard star was used to obtain the photometric calibration and the atmospheric extinction parameters. We also carried out spectroscopic observations of two galaxies, namely KUG~1013+381 and I~SZ~59, in our sample using the 2-meter Himalayan Chandra Telescope (HCT), Hanle which is operated by the Indian Institute of Astrophysics (IIA), Bangalore, India. The details of the spectroscopy observations will be presented elsewhere. Here, the results from these spectroscopy observations are used to apply the internal reddening corrections to the \Ha flux for these two galaxies.

\section{Data Reduction and Calibration}

\subsection{Image processing}

The CCD images were cleaned following the standard procedures of bias subtraction and flat-fielding using the CCDPROC task of IRAF (Image Reduction and Analysis facility) software developed by {\it National Optical Astronomy Observatory (NOAO)}. The dark currents in the CCDs used are negligible and hence no dark subtraction was applied. The cosmic rays were removed interactively using the IRAF tasks COSMICRAYS and CREDIT within the CRUTIL package. The observed frames for a galaxy were aligned using the IRAF tasks GEOMAP and GEOTRAN. As the image frames were taken at different values of airmass, the images were corrected for the atmospheric extinction before combining the aligned frames. The noise weighted mean of these frames was estimated to get the combined frame for a particular filter. The weighting factor multiplied to each frame was $1/\sigma^2$, where $\sigma$ is the background sky rms for the frame.

\subsection{Continuum subtraction}

The \Ha emission-line flux measured by the narrow-band filter includes both the line flux and the continuum flux. This continuum emission is dominated by stars in the galaxy. It is therefore necessary to subtract the continuum in order to estimate the \Ha emission-line flux. The underlying continuum in the \Ha filter was subtracted using the standard procedure given by Waller (1990) and Spector et al.\ (2012). In this method, the underlying continuum emission is estimated from the nearest wide-band such as $r$-band filter, which is significantly wider compared to the narrow band. The main assumption here is that the galaxy continuum flux per unit wavelength does not differ much across the narrow-band and the wide-band filters. The continuum emission in the wide-band filter is scaled to that in the narrow-band filter by using a scaling factor, called the Wide to Narrow Continuum Ratio (WNCR), defined as the ratio of the continuum count rate in the wide-band filter to the continuum count rate in the narrow-band filter. The count rates are determined using unsaturated, non-variable, isolated field stars in the wide and narrow band images. Here, it is important that the field stars should not have any strong emission or absorption line within the wide-band and narrow-band filters. The average WNCR for each set of the \Ha and $r$-band filters is given in Table~\ref{cal}. The WNCR value is used to scale down the $r$-band image so that the counts per second of field stars in the $r$-band and the H$\alpha$-band are almost identical. The continuum subtracted \Ha images were obtained by subtracting the scaled $r$-band images from the H$\alpha$-band images. After continuum subtraction, all field stars should normally be completely removed from the image. However, in practice, some residual emission may be seen at the locations of bright stars due to imperfections in the assumptions made and some variations in the seeing conditions during the observations. Our observations in the \Ha and $r$-bands were always taken during the same night, which minimized effects due to extreme variations in inter-night seeing and in most of the cases, the continuum subtraction process worked very well. Only in a few frames, some residual continuum emission was visible at locations of very bright stars in the frame. The typical error in the estimates of WNCR values is 5\%.  

\subsection{Astrometry}

The images were registered for the equatorial coordinates in the J2000 epoch using the {\it Two Micron All Sky Survey} (2MASS) or SDSS $r$-band images of the same region. In this process, the World Coordinate System (WCS) solutions are calculated using the pixel coordinates of a set of stars in the target frame and the WCS coordinates of the same stars in the reference frame using the IRAF task CCMAP. On applying the WCS solutions on the target frames, the astrometric uncertainty is found to be less than $0.3''$. The same astrometric calibration was applied to the continuum subtracted \Ha image. These \Ha and $r$-band images were made in publishable format using IGI (Interactive Graphics Interpreter) package of STSDAS in IRAF.

\subsection{\Ha flux determination and calibration}

In order to obtain integrated \Ha flux, the polygon aperture photometry was performed on the continuum subtracted \Ha images using the IRAF task POLYPHOT. The size of the aperture was selected such that it included all the star-forming regions as well as diffuse \Ha emission from the galaxy. We have not attempted here to measure \Ha flux for individual star forming regions and diffuse \Ha emission separately. The average sky background value per pixel was measured close to the galaxy. The \Ha flux was calibrated using observations of spectrophotometric standard stars taken at different values of airmass. The atmospheric extinction was estimated from the observations of these standard stars. The flux (counts~s$^{-1}$) of the standard star was estimated using the DAOPHOT package in IRAF. The expected \Ha flux (erg~s$^{-1}$cm$^{-2}$) of the standard star was calculated by integrating the product of the calibrated spectrum of standard star and the filter transmission curve with respect to the wavelength. The instrumental response was obtained by dividing the expected \Ha flux (erg~s$^{-1}$cm$^{-2}$) of the standard star with its observed \Ha flux (counts~s$^{-1}$) at zero airmass. The instrumental responses for the combinations of telescopes and filters are given in Table~\ref{cal}. The calibrated \Ha flux for a galaxy was estimated by multiplying its observed \Ha flux with the instrumental response. The \Ha flux needs to be further corrected for an attenuation in the \Ha filter transmission due to the shift in the emitted wavelength of the \Ha emission line at the redshift of the galaxy. The narrow-band filters normally have maximum transmission at the rest frame wavelength of the \Ha emission. The narrow-band filter response curve is also dependent on the focal ratio of the telescope and requires a correction. The effect of the telescope focal ratio is important only for the 1.3-meter DFOT, since the filters are placed here in the fast (f/4) converging beam. Both of these corrections were estimated in Jaiswal \& Omar (2013) and were applied here. 

The calibrated \Ha flux was corrected for line contaminations in the filter passbands, the Galactic extinction and the internal extinction. The narrowband (H$\alpha$) filters contain significant flux from [\NII]~$\lambda\lambda$6548,6584 lines. The wide-band ($r$) filter, in addition, will contain significant flux from [\OI]~$\lambda$6300, [\NII]~$\lambda$6548, [\NII]~$\lambda$6584, [\SII]~$\lambda$6717 and [\SII]~$\lambda$6731 emission lines apart from the \Ha$\lambda$6563 line. It should be noted that the \Ha line is a contaminating line in the $r$ band for estimating continuum flux. The \Ha line flux should be corrected for these line contaminations. The fluxes of the emission lines contaminating the filter passbands were estimated from the SDSS spectroscopic data. It is assumed here that the line ratios are constant everywhere in the galaxy. The galactic (foreground) extinction correction was determined using $E_f(B-V)$ values based on Schlafly \& Finkbeiner (2011) recalibration of the Schlegel, Finkbeiner \& Davis (1998) extinction map. These values were taken directly from NED. Assuming an intrinsic Case-B recombination ratio of 2.86 (Osterbrock 1989) for H$\alpha$/H$\beta$ ratio, valid for an ionized gas with an electron temperature of 10$^4$K and an electron density of 100~cm$^{-3}$), and the Cardelli, Clayton \& Mathis (1989) extinction curve with $R_V= 3.1$, the internal reddening of the ionized gas can be expressed in terms of the observed H$\alpha$/H$\beta$ flux ratio (corrected for the Galactic extinction) as: 

\begin{equation}
E_g(B-V) = 2.33 \times \mathrm{log} \left(\frac{f_{H\alpha}/f_{H\beta}}{2.86}\right)
\end{equation}

The values of $E_f(B-V)$ and $E_g(B-V)$ for each galaxy are listed in Table~\ref{archive}. The $E(B-V)$ values can be changed into the extinction coefficients at the \Ha wavelength using the Cardelli, Clayton \& Mathis (1989) extinction curve with $R_V= 3.1$ as: $A_{H\alpha} = 2.54 ~E(B-V)$. The H$\alpha$ and H$\beta$ fluxes for several galaxies are taken from the MPA-JHU (Max-Planck-Institute for Astrophysics and Johns Hopkins University) emission line analysis database for the SDSS data release-7 (DR7), which were already corrected for the foreground extinction using O'Donnell (1994) extinction law. Therefore, the $E(B-V)$ value calculated using these line fluxes will only provide the internal extinction without a contribution from foreground extinction due to the Milky-way. Therefore, we applied the foreground reddening corrections separately for those galaxies, for which MPA-JHU databases were used. The ratio for the galaxies MRK~475, UGCA~116 and UGCA~130 were taken from the literature. We estimated the H$\alpha$/H$\beta$ flux ratio for galaxies I~SZ~59 and KUG~1013+381 from our spectroscopic observations. The H$\alpha$/H$\beta$ flux ratios for these galaxies (MRK~475, UGCA~116, UGCA~130, I~SZ~59 and KUG~1013+381) provide sum of foreground and internal extinctions.

\subsection{Errors in \Ha flux estimates}

The accuracy of \Ha flux estimates from the narrow-band \Ha imaging depends on the accurate removal of the underlying continuum. The error in WNCR gives main contribution to the total error (Spector et al. 2012). We used $10 - 15$ foreground stars to calculate the WNCR value. The residual flux at the locations of bright field stars after subtraction of the scaled continuum is similar to error in WNCR (below 5\%), the errors in continuum subtraction for galaxies can be fixed at this value. In some cases (UM~311, NGC~941, NGC~1087 and UGCA~130), flat-fielding was not perfect and a large-scale residual gradient was seen in the cleaned \Ha images. However, we always estimated local sky background around galaxies, in order to minimize effects of improper flat-fielding on flux estimates. These effects are expected to be insignificant compared to other errors. In addition to the WNCR error, there will be some additional sources of observational errors arising from uncertainties in the estimates of instrumental response, narrow-band filter transmission curve and atmospheric extinction. It is difficult to make an exact estimate for errors in the \Ha flux in individual cases. We believe that typical errors in \Ha flux corrections remain below 10\%. The final errors in the \Ha flux estimates are typically between 10\% and 20\%.

\section{Archival Data}

The star formation rates in galaxies can be well constrained by multi-wavelength data. In the present study, we have used the archival data at FUV, FIR and 1.4~GHz radio bands to make independent estimates for SFR. The FUV data is taken from the $Galaxy~Evolution~Explorer$ (GALEX) survey. The details of the GALEX telescope, detectors and data products are given in Morrissey et al. (2007). The GALEX telescope has two photometric bands, FUV (1344-1786~$\rm \mathring{A}$; $\lambda_{eff} = 1528 ~\rm \mathring{A}$) and NUV (1771-2831~$\rm \mathring{A}$; $\lambda_{eff} = 2310 ~\rm \mathring{A}$). We estimated FUV flux densities for galaxies in our sample from the GALEX FUV images using POLYPHOT task of IRAF. The measured counts per second ($CPS$) in the FUV image is converted into flux density using the calibration relation: $f_{FUV}({\rm erg~s^{-1}~cm^{-2}~\mathring{A}^{-1}}) = 1.40 \times 10^{-15} \times CPS$. This flux density is corrected for the Galactic extinction using $E_f(B - V)$ values in Table~\ref{archive} and for the internal extinction of the stars using $E_s(B - V)$ values. Following Calzetti (1997), the color excess of the stellar continuum $E_s(B - V)$ is related to the color excess of the ionized gas $E_g(B - V)$ as: $E_s(B - V) = (0.44 \pm 0.03) ~E_g(B - V)$. The total $E(B - V)$ will give the extinction $A_{FUV} = 8.15 ~E(B - V)$ using the Cardelli, Clayton \& Mathis (1989) extinction curve with $R_V= 3.1$. The corrected FUV flux densities for the sample galaxies are given in Table~\ref{flux_a}.

The FIR data is taken from the IRAS survey (Neugebauer et al. 1984). The IRAS mission performed a low-resolution all sky survey at 12 $\mu$m, 25 $\mu$m, 60 $\mu$m and 100 $\mu$m. We have used the 60 $\mu$m and the 100 $\mu$m FIR flux densities as given by NED to get an estimate of FIR emission from the sample galaxies. These flux densities are given in Table~\ref{flux_a}. The 1.4~GHz radio continuum data is taken from the FIRST and NVSS databases. The FIRST survey (Becker, White \& Helfand 1995) images have a typical rms of 0.15 mJy beam$^{-1}$ and an angular resolution of 5$''$. On the other hand, NVSS (Condon et al. 1998) images have a typical rms of 0.45 mJy beam$^{-1}$ and a resolution of 45$''$. The NVSS images are more sensitive to extended emission. We used the FIRST data for small galaxies and the NVSS data for large galaxies. The 1.4~GHz flux densities from the NVSS are taken directly from NED. We convolved the FIRST images with $20''\times 20''$ beam size to make image resolution comparable to the source size, before estimating flux densities. The AIPS (Astronomical Image Processing System; Greisen 2003) developed by the National Radio Astronomy Observatory (NRAO) was used for analysis of  radio images.

\section{Results}

The grey-scale \Ha and $r$-band images of all the galaxies are presented in Figure~\ref{images}. A linear scale in kilo-parsec is shown at the bottom of the each image. The \Ha sensitivities of the images presented here are typically $\sim$ 10$^{-16}$ erg~s$^{-1}$~cm$^{-2}$~arcsec$^{-2}$. The observed sample was conveniently divided into two major classes: large spiral galaxies and dwarf or small galaxies. The large galaxies include NGC~1087, NGC~2799, NGC~3294, NGC~3381, NGC~3423, NGC~3430, NGC~3949, NGC~4389 and NGC~0941. On the other hand, the dwarf or small galaxies include CGCG~041-023, IC~0225, IC~2524, IC~2828, IC~3521, I~SZ~059, KUG~1013+381, MRK~0022, MRK~0475, SBS~1222+614, UGC~09273, UGCA~116, UGCA~130, UM~311 and UM~439. The \Ha flux estimates for all the galaxies of our sample are provided in Table~\ref{flux}. The second column ($F_{H\!\alpha}$) of this table lists the calibrated but un-corrected \Ha flux. The subsequent columns show corrected \Ha flux in sequential steps as: (i) $F^i_{H\!\alpha}$ for line (e.g., \NII~6548 and \NII~6584) contamination in the narrow-band \Ha filter, (ii)  $F^{ii}_{H\!\alpha}$  for line (e.g., [\OI]~6300, [\NII]~6548, \Ha~6563, [\NII]~6584, [\SII]~6717 and [\SII]~6731) contamination in the wide-band $r$ filter, and finally, (iii) $F^{iii}_{H\!\alpha}$ was obtained by correcting $F^{ii}_{H\!\alpha}$ for the galactic extinction and the internal galaxy extinction using $E(B-V)$ values.

\subsection{Star formation rates}

The star formation rates in galaxies can be estimated using fluxes at different wavelengths, viz., FUV, optical, IR and radio (e.g., Kennicutt 1998a; Buat et al. 2002; Rosa-Gonz{\'a}lez et al. 2002; Hopkins et al. 2003; Murphy 2011). The reliability of each of these indicators has been widely discussed (e.g., Condon 1992; Kennicutt 1998a; Calzetti et al. 2007; Salim et al. 2007). The complexity of the astrophysics underlying each method leads to a considerable degree of uncertainty in its use as an estimator of the star formation rate. Therefore, it is useful to estimate SFR using all possible methods. The SFR is calculated by multiplying luminosity in a wave-band by a calibrated numerical factor. The numerical factor for each band is computed using stellar evolutionary synthesis models and certain assumptions on the physical processes in the ISM. There are slight differences in the conversion factors for a wave-band investigated in different research works. These minor differences arise due to usage of different stellar evolution models, atmosphere models, metal abundances, initial mass functions (IMF) and star formation histories. Such variations have been discussed previously (e.g., Kennicutt 1998a; Murphy et al. 2011). For instance, the conversion factor between SFR and \Ha luminosity given by Calzetti et al. (2007) is 0.67 times of that given by Kennicutt (1998b). The calibration constant is also dependent on the age of the most recent star formation event and star formation history in galaxies. Over a time scale of 100 Myr, the differential changes in the calibration constants are small. However, such effects may be very important for young star-bursts like WR galaxies. A detailed description of conversion factor and its sensitivity to different parameters is beyond the scope of this paper. A brief overview of methods used to convert luminosities to SFR is provided in the Appendix. The luminosities at different wave-bands and the corresponding estimates of SFRs are presented in Table~\ref{sfr}. This table also lists an assumed SFR estimated as a median value, obtained from all the estimates of SFR.

\subsection{\Ha and $r$-band morphologies}

\begin{figure*}
\centering
\epsfig{figure=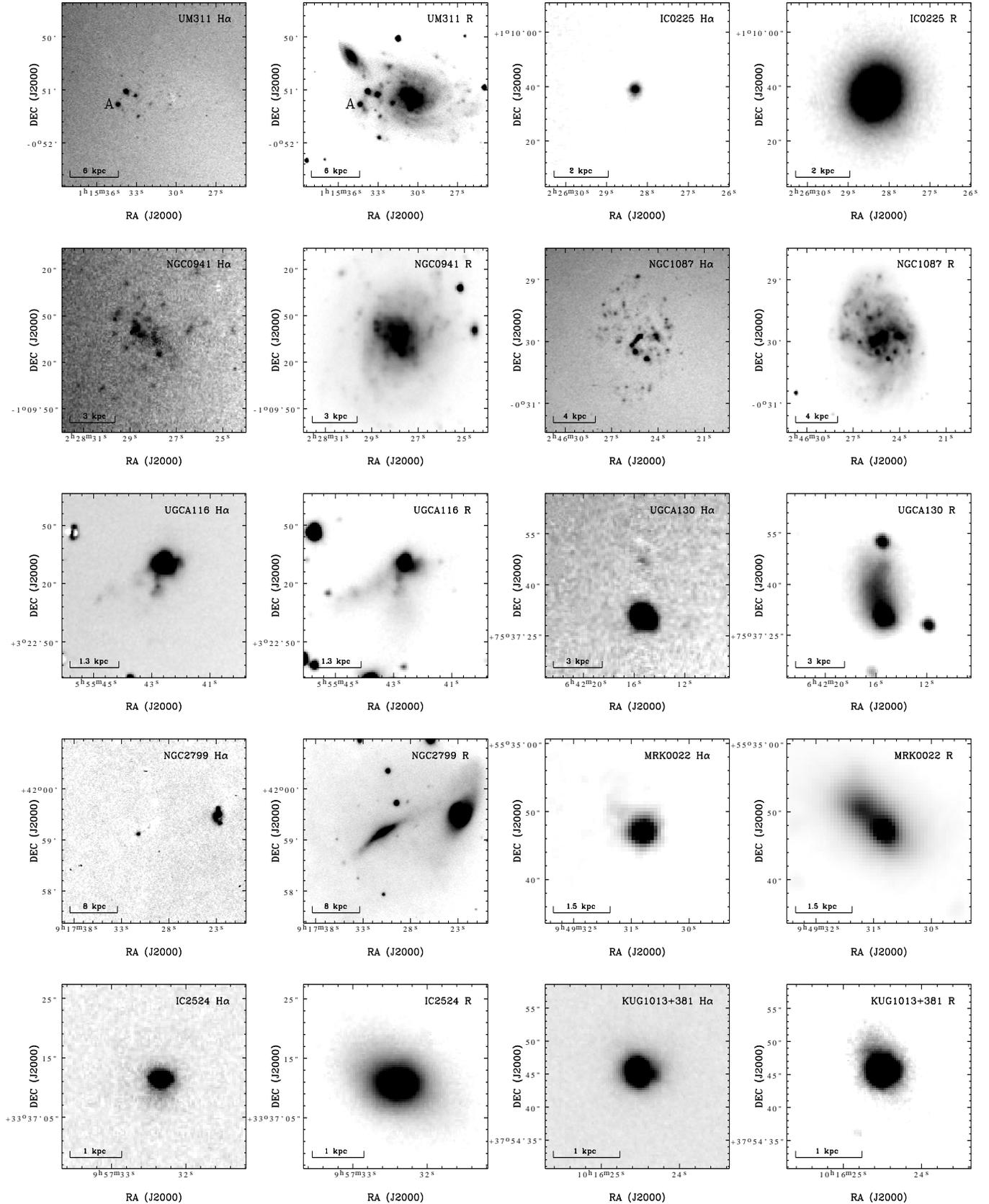,scale=0.92,angle=0}
\vspace{-1cm}
\caption{\Ha and $r$-band images of the WR galaxies in our sample. North is up and East is to the left. A linear scale-length in kpc is shown at the bottom of each image.}
\label{images}
\end{figure*}

The following notes summarize morphological features (see Figure~\ref{sdss} and Figure~\ref{images}) seen for each galaxy in our sample:\\

\hspace{-0.64cm}{\bf UM~311}\\
UM~311 (shown by label `A' in Figure~\ref{images}) is considered as a giant \HII region complex in the spiral galaxy NGC~450 or a dwarf \HII galaxy interacting with NGC~450. The optical observations has also been presented in Karthick et al. (2014). It has not been possible to favour any one scenario over the other based on optical observations. Another smaller spiral galaxy UGC~807 in the field is a background galaxy, and does not form a physical pair with NGC~450. We looked at high resolution 1.4~GHz image from the FIRST survey and noticed that the radio emission is resolved primarily into four knots. The radio emission is not detected from the position of UM~311, but detected from two other nearby knots, which are also detected in H$\alpha$. The FIR emitting regions seen in the Herschel images are coincident with the radio emitting knots. We also notice that this \HII region complex is in a ring shape. In SDSS images, this complex is the bluest among other \HII regions in the galaxy. It leaves us with an interesting possibility that UM~311 may be a dwarf galaxy interacting with NGC~450.\\

\hspace{-0.64cm}{\bf IC~225}\\
IC~225 is classified as a dwarf elliptical (dE) galaxy. The broad-band SDSS optical images revealed two nuclei separated by 1.4$''$ - a central core along with a blue region, which disappears in the redder bands (Gu et al. 2006). The \Ha emission is seen centrally concentrated in our images, and the nucleus is not resolved. No radio emission is detected from this galaxy. \\

\hspace{-0.64cm}{\bf NGC~941}\\
NGC~941 is a barred spiral galaxy with multiple spiral arms. It forms a non-interacting pair with NGC~936, at 12.6$'$ separation (de Vaucouleurs et al. 1976). The star forming \Ha emitting regions are seen spread over the disk with a lopsided central concentration of bright \HII regions with a bar like feature.\\

\hspace{-0.64cm}{\bf NGC~1087}\\
NGC~1087 is a face-on large spiral galaxy with multiple spiral arms and a small bar. It is a member of NGC~1068 Group (de Vaucouleurs et al. 1976) and has a close companion galaxy NGC~1090. Blackman (1980) considered NGC~1087 as an isolated galaxy due to the large velocity difference between NGC~1087 ($v=1517~\rm km~s^{-1}$) and NGC~1090 ($v=2760~\rm km~s^{-1}$). Martin \& Friedli (1997) quoted two asymmetries in this galaxy, namely a large (13\deg) misalignment between the \Ha bar and the stellar bar, and more star formation regions in the north, compared to the south. Hummel et al. (1987) found extended 1.4 GHz radio emission with nearly $1'.1$ offset from the optical center. The \Ha emission is seen from the nuclear bar. Several bright \HII regions are also seen along the spiral arms. This galaxy is highly lopsided.\\

\hspace{-0.64cm}{\bf UGCA~116}\\
The BCD galaxy UGCA~116 is a cometary type galaxy having a `head-tail' structure (e.g., Cair{\'o}s et al. 2001). Such an unusual morphology is expected from a merger between two small galaxies (Baldwin, Spinrad \& Terlevich 1982; Brinks \& Klein 1988). Optical and radio observations (e.g., Sage et al. 1992; Deeg et al. 1997; van Zee et al. 1998) shows an extraordinary star formation, which indicates that UGCA~116 might be observed at the peak of its star forming episode caused by a rare merger between two gas-rich dwarf galaxies. Bright \Ha emission is seen from the head region along with some relatively fainter \Ha knots and \Ha arcs along two tails.\\

\hspace{-0.64cm}{\bf UGCA~130}\\
The BCD galaxy UGCA~130 is also a cometary type galaxy having a `head' towards the south and a `tail' extending towards the north with a bright source at the end. The \Ha emission is seen mainly in the head of the cometary type structure. Faint \Ha emission is also detected from the bright source towards the end of the tail. L{\'o}pez-S{\'a}nchez (2010) speculated that UGCA~130 is not undergoing any recent interaction but it has a disturbed \HI kinematics due to some past interaction.\\

\hspace{-0.64cm}{\bf NGC~2799}\\ 
NGC~2799 is a member of close interacting galaxy pair with NGC~2798. It is an edge-on galaxy with a prominent bar. A common stellar envelop fills the system indicating that two galaxies are in an advanced stage of interaction leading to a possible merger. Both galaxies show prominent star formation in the \Ha images. NGC~2798 is classified as a starburst galaxy (Kinney et al. 1993). Several extra-planner \Ha emitting knots are detected in the common stellar envelop. These \Ha knots may be tidal dwarf galaxies or \HII regions in the disturbed stellar envelope caused by tidal interaction. \\

\hspace{-0.64cm}{\bf MRK~22}\\
MRK~22 is a blue compact galaxy having double nuclei (Mazzarella et al. 1991). A strong \Ha emission is seen from the bright nucleus. The double nuclei nature of this galaxy indicates that it is likely a merger of two dwarf systems (Pustilnik et al. 2001). A diffuse \Ha emission tail is seen extending to the other nucleus.\\

\hspace{-0.64cm}{\bf IC~2524}\\
IC~2524 is classified as a dwarf disk galaxy. The disk is very faint. This galaxy has no visible companion within 30$'$ radius. The \Ha emission is seen in the nuclear region along with weak diffuse emission surrounding the nucleus.\\

\hspace{-0.64cm}{\bf KUG~1013+381}\\
KUG~1013+381 is a blue compact dwarf galaxy having intense nuclear star formation. A faint small diffuse region is seen in $r$-band image at $\sim9''$ towards the south. Pustilnik \& Martin (2007) speculated that the recent starburst phase in this galaxy has been triggered through the interaction of its binary companion (UGC~5540) at an angular separation of $\sim8'$. The \Ha image shows bright nuclear region with two \Ha emission regions.\\

\hspace{-0.64cm}{\bf NGC~3294}\\
NGC~3294 is classified as a SA(S)c type galaxy. The WR features in this galaxy are detected from a very bright cemetery type region located in the west, in between the two spiral arms. This region is the brightest in H$\alpha$. The color of this region is significantly bluer than the galaxy. This region has a measured recession velocity as 1464 $\rm km~s^{-1}$ (Abazajian et al. 2009) compared to the galaxy's recession velocity as 1586 $\rm km~s^{-1}$ (de Vaucouleurs et al. 1991). We speculate that this WR \HII region is a dwarf galaxy undergoing a minor interaction with NGC~3294.\\

\hspace{-0.64cm}{\bf NGC~3381}\\
NGC~3381 is a spiral galaxy with two prominent spiral arms and a bright bar. The galaxy is highly lopsided in its stellar light distribution having diffuse envelope towards the west. The star formation is seen mainly in the nuclear region. Faint \Ha emission is also seen in small patches along the spiral arms.\\

\hspace{-0.64cm}{\bf NGC~3423}\\
NGC~3423 is a face-on spiral galaxy with multiple spiral arms and a bright nucleus. A very high concentration of star forming regions is seen in the disk. The \Ha region density is high in the central region. Ganda et al. (2007) noticed that the star-burst in this galaxy is very recent on the basis of their absorption line strength analysis. Condon (1987) found a disturbed radio morphology at 1.4~GHz radio band.\\

\hspace{-0.64cm}{\bf NGC~3430}\\
The spiral galaxy NGC~3430 forms a pair with a star forming spiral galaxy NGC~3424 at about 6$'$ (36 kpc) separation with 100 \kms ~velocity difference (Braine et al. 1993). Nordgren et al. (1997) found a sign of tidal interaction in both the galaxies through optical and \HI imaging. The star forming regions are seen in the nucleus and spiral arms.\\

\hspace{-0.64cm}{\bf CGCG~038-051}\\
CGCG~038-051 is a dwarf \HII galaxy of irregular shape. Three prominent star forming knots are identified along the major axis of the galaxy in the \Ha image. The presence of multiple distinct nuclei within an asymmetric diffuse stellar envelope suggests that this system is undergoing a merger.\\

\hspace{-0.64cm}{\bf IC~2828}\\
IC~2828 has an irregular morphology. The \Ha image shows multiple bright \Ha regions, tails and plumes like features. Two bright \Ha regions are identified at the outer edges of this galaxy. The optical $r$-band image shows a faint tail extending towards the south. These features are suggestive of a recent tidal interaction or a merger event.\\

\hspace{-0.64cm}{\bf UM~439}\\
UM~439 has an irregular morphology with two prominent nuclei. The \Ha image reveals multiple nuclei. The tail and plume type features are also seen in the galaxy. Taylor et al. (1995) have found an asymmetric \HI distribution, but they have not found any visible companion. These features are suggestive of a recent tidal interaction or a merger event.\\

\hspace{-0.64cm}{\bf NGC~3949}\\
NGC~3949 is a spiral galaxy with multiple spiral arms and a small bar. It is a member of the Ursa-Major cluster of galaxies. Maslowski \& Kellermann (1988) mapped this galaxy at 5~GHz radio band using VLA `B' configuration and found that the location of the peak radio intensity has an offset of about $12''$ from the optical nucleus of the galaxy. Several star forming knots are identified in this galaxy, typical of a normal spiral galaxy.\\

\hspace{-0.64cm}{\bf I~SZ~059}\\
Gil de Paz, Madore \& Pevunova (2003) classified this object as nE BCD galaxy. This galaxy has a very elongated elliptical envelope within which a clearly defined nucleus exists. Doublier, Caulet \& Comte (1999) have shown using surface brightness distribution that it has a disk. This galaxy is a member of the loose group NGC~4038 (Firth et al. 2006). Strong \Ha emission is seen from the nucleus. The $r$-band image reveals diffuse stellar envelop.\\

\hspace{-0.64cm}{\bf CGCG~041-023}\\
CGCG~041-023 is a barred spiral galaxy. Two symmetrically located prominent star forming regions are identified on the edges of the galaxy. This morphology is very intriguing. A weak \Ha emission is also seen along the bar. Further observations are required to understand the observed morphology of this galaxy.  \\

\hspace{-0.64cm}{\bf SBS~1222+614}\\
SBS~1222+614 is a dwarf irregular \HII galaxy. The diffuse \Ha emission is seen in the galaxy. A weak diffuse \Ha emitting arc is also visible in the south-east direction of the galaxy. \\

\hspace{-0.64cm}{\bf NGC~4389}\\
NGC~4389 is a barred spiral galaxy. Most of the star formation is concentrated along the bar. Some \Ha emitting knots are also seen in the disk.  According to Sandage \& Bedke (1994), the spiral arms of NGC~4389 are difficult to trace because of the large inclination angle. A few isolated extra-planner \Ha emitting regions are detected in the halo of the galaxy. It is likely that this galaxy has undergone an interaction in the recent past. \\

\hspace{-0.64cm}{\bf IC~3521}\\
IC~3521 has an irregular morphology with a bar. This galaxy is a member of the Virgo cluster. Two prominent star forming knots are identified at the edges of the galaxy.  The $r$-band morphology is patchy with a common diffuse stellar envelope. The stellar light distribution suggests an ongoing merger.  \\

\hspace{-0.64cm}{\bf UGC~9273}\\
UGC~09273 is classified as an irregular galaxy. The multiple star forming knots are detected in a ring like structure, which is very intriguing. The stellar light distribution is lopsided. \\

\hspace{-0.64cm}{\bf MRK~475}\\
MRK~475 is classified as a BCD galaxy. It shows strong nuclear star formation. The $r$-band morphology indicates lopsidedness in this galaxy. \\

\section{Discussions}
\subsection{Comparison of different SFRs}

The estimated values of SFRs using the luminosities in different wave-bands are given in Table~\ref{sfr}. The Figure~\ref{sfr1} shows comparisons of the H$\alpha$ based SFR with the SFRs derived using the FUV, FIR and 1.4~GHz radio luminosities. In these plots, we have also included SFR estimates of WR galaxies provided in L\'opez-S\'anchez (2010) in order to enlarge our sample size. The small and large spiral galaxies are designated by filled and open circles respectively. The error-weighted linear fits (solid line) to the data points were made. It can be seen from Figure~\ref{sfr1} that our sample is dominated by galaxies with SFR $\le 1 M_\odot~{\rm yr}^{-1}$, while majority of galaxies in the L\'opez-S\'anchez (2010) sample have SFR $\ge 1 M_\odot~{\rm yr}^{-1}$. These two samples are therefore complimentary to each other. The H$\alpha$-based SFR is seen well correlated with other SFR values estimated using FUV, FIR and 1.4~GHz radio luminosities in these plots. The strongest correlation is seen between \Ha and FUV SFRs. The slope of the H$\alpha$-FUV correlation is close to unity. This result is in agreement with previous studies (e.g., Buat, Donas \& Deharveng 1987; Buat 1992; Sullivan et al. 2000; Bell and Kennicutt 2001; L\'opez-S\'anchez 2010). The slopes of the FIR-\Ha and radio-\Ha SFR relations are $\sim 1.3$ and $\sim 1.5$ respectively. The correlation coefficient and scatter for FUV-\Ha and FIR-\Ha are almost identical at $\sim 0.9$ and  $\sim0.4$. The radio-\Ha correlation shows larger scatter compared to other correlations. 

Various interplay between the calibration constants and age, IMF, and other physical processes in ISM are described in Calzetti (2012). The tight correlation between FUV and \Ha SFR is not surprising. The \Ha based SFR estimates trace the youngest star formation. The FUV photons from the most massive stars ($M \ge 20~M_\odot$) are responsible for the \Ha line. The ionizing FUV flux decreases by two orders of magnitude between 5 Myr and 10 Myr after the burst. While the massive O-type stars die in a few tens of Myr time scale, the out-numbered B-type stars continue contributing to the total FUV flux over longer time scales. On shorter time scales of a few Myr, the total FUV flux emitted by O and B type stars collectively can vary up to a factor of 3. The calibration constants for converting luminosities to SFR in different bands are only an approximation. The best accuracies in these calibrations are $\pm15$\%. The internal extinction can be quite significant in the FUV band, but relatively less significant in the \Ha band. It also appears from Figure~\ref{sfr1} that scatter is slightly larger towards the low SFR regime, which is dominated by small-size galaxies. The log-log rms scatter values between 0.4 and 0.5 in these correlations can be understood within various uncertainties and inaccuracies in calibrations, particularly for WR galaxies where star-burst is very young. Our results are largely consistent with previous studies in terms of scatter in various SFR correlations (Bell and Kennicutt 2001; Lee et al. 2009; Hao et al. 2011). In small galaxies, the most recent single burst dominated by WR region may often be representing the total star formation. On the other hand, in large galaxies, episodic star formation spread over longer time scales (a few 100 Myr) may be present. This conjecture is supported by the color composite SDSS and \Ha images, which show that large galaxies have wide-spread star forming disks, often with a distinct very blue star forming region. These blue regions are often identified as the WR regions in the galaxy. The total SFR as measured here in large galaxies are therefore expected to be an average over longer time period. In the present study, we have considered only global star formation rates in galaxies. Therefore, large galaxies in the sample are not tracing distinctly the emission in WR regions alone. It will be interesting to examine local SFR correlations in the WR galaxies. 

The WR galaxies are also showing a tight correlation between FIR  and \Ha SFRs. The FIR luminosity was determined using 60 $\mu$m and 100 $\mu$m flux densities. The slope of the FIR-\Ha SFR correlation is significantly deviating from unity with an indication that FIR fluxes are under-estimating SFR in the lower SFR sides. The FIR emission is thermal emission from the dust heated by FUV photons, emitted by hot and young stars. The FIR luminosity depends also on the dust content and its distribution relative to star forming regions. The calibration for FIR luminosity to SFR is largely based on large galaxies where older stars also contribute significantly to the heating of dust. We speculate that in low SFR range, star-light spectral energy distribution (SED) plays a more deciding role for 60 $\mu$m and 100 $\mu$m fluxes. A total (bolometric) infrared luminosity may give better estimates for SFR (Calzetti 2013). Finally, the radio and \Ha SFRs are also correlated, but with the largest scatter. The radio emission is primarily of synchrotron (non-thermal) origin in star forming galaxies, with relatively small contribution from thermal free-free emission from the ionized gas. The slope of the radio-\Ha SFR correlation indicates that towards high SFR ends, radio emission is consistent with that expected from the \Ha emission. However, radio emission is under-estimating SFR towards the low SFR range. The scatter is also significantly higher towards the low SFR range. The non-thermal radio continuum emission is expected to last for longer duration, nearly 100 Myr after the initial starburst, compared to the \Ha emission. The episodic star-bursts separated by a few tens of Myr in a galaxy will further complicate a comparison between these tracers of SFR. There is a possibility that young supernovae responsible for accelerating cosmic electrons to the relativistic speeds are lacking in the very young starbursts in WR galaxies. Several dwarf galaxies in our sample are not detected in the radio continuum. The radio emission from galaxies also depends on the strengths of magnetic field in galaxies. The magnetic field in dwarf galaxies is still poorly understood (Beck \& Wielebinski  2013). 

In order to further understand radio emission from WR galaxies, we estimated expected thermal radio continuum from the knowledge of the \Ha flux in these galaxies, using the relation given by Dopita et al. (2002), 

\begin{equation}
F_{1.4GHz (thermal)} ~[\rm mJy] = 1.21 \times 10^{21} ~F_{H\alpha}~[\rm erg~cm^{-2}~s^{-1}]
\end{equation}

The thermal radio continuum is subtracted from the total radio emission to get an estimate for non-thermal radio continuum. The ratio (R) of non-thermal to thermal radio flux is given in Table~\ref{R}.  The average value of this ratio for a sample of star-burst galaxies was estimated as log R = $1.3 \pm 0.4$ (Dopita et al. 2002). It can be seen that several WR galaxies in our sample have significantly lower values of R. This is indicating a lack of synchrotron emission in several WR galaxies. This ratio is plotted in Figure~\ref{rplot} against the ratio of SFR(FUV) and SFR(H$\alpha$). This plot indicates a weak positive correlation between these two quantities. The higher values of FUV to \Ha ratio are expected in relatively older star-bursts as the \Ha emission decreases faster than total FUV flux in the first tens of Myr from the initial burst time. In older star-bursts, supernovae explosions are expected to give rise to normal non-thermal radio flux. Therefore, this hint of correlation between `R' and SFR(FUV)/SFR(H$\alpha$) supports our initial observation that majority of WR galaxies are lacking non-thermal radio flux. 

\subsection{Radio-FIR correlation}

A tight correlation between radio continuum and FIR emission is known in normal star forming galaxies (e.g., Helou et al. 1985; Condon 1992; Niklas \& Beck 1997; Yun et al. 2001). The radio-FIR correlation in star-forming galaxies is understood as follows. The dust absorbs FUV photons produced by hot stars, and thereby gets heated. This warm dust re-radiates in the FIR waveband, producing a linear correlation between SFR and FIR luminosity. The GHz radio continuum emission from star-forming galaxies is mainly synchrotron radiation from the cosmic electrons accelerated to relativistic speeds through supernova explosions of the young massive stars. Therefore, a radio-FIR correlation is expected in star forming galaxies. The tightness of the radio-FIR correlation is estimated in terms of a parameter ($q$), which is defined as the logarithmic of the ratio of total FIR flux density and radio continuum flux density at 1.4 GHz (see, Helou et al. 1985).

\begin{equation}
q = {\rm log} \left (\frac{2.58~S_{{60\mu {\rm m}}} + S_{{100\mu {\rm m}}}}{2.98~S_{1.4GHz}} \right)
\end{equation}

\noindent where all flux densities are in Jy. The average value of $q$ for normal galaxies has been estimated as $<\!q\!>\sim2.34 \pm 0.26$ (Yun et al. 2001). The most remarkable feature of the radio-FIR correlation is that it displays a very tight correlation in galaxies spanning 5 orders of magnitude in luminosity (Price \& Duric 1992). 

The comparison of the FIR based SFR with the radio based SFR, and the radio-FIR correlation for the WR galaxies in the present sample are shown in Figure~\ref{correlation}. At high SFR values, we find that the SFR estimates using the 1.4 GHz luminosity are consistent with that using the FIR luminosity. However, the radio emission underestimates SFR towards the low SFR range. The slope for the SFR(radio) versus SFR(FIR) is $\sim$1.3. The $q$ values are falling within the scatter seen for normal galaxies. However, we notice a possible trend that galaxies at low FIR or low SFR end, are showing higher value of $q$. These trends can again be understood as a general deficiency of non-thermal radio flux in WR galaxies as discussed in the previous section. It is worth to point out that some WR galaxies such as MRK~996 (Jaiswal \& Omar 2013), and several galaxies in L\'opez-S\'anchez (2010) sample show N/O enrichment. The N/O enrichment is expected to take place in young (WR) star-burst regions, where nitrogen is enriching the ISM through the stellar winds, however, similar oxygen enrichment expected from supernovae explosions is lacking (Kobulnicky et al. 1997; Putilnik et al. 2004; Brinchmann et al. 2008; L\'opez-S\'anchez \& Esteban et al. 2010b). Although, magnetic field strengths in dwarf galaxies are not known very well, a few radio polarimetric observations of dwarf galaxies reveal presence of magnetic fields in these systems. (Chy{\.z}y et al. 2000; Gaensler et al. 2005; Chy{\.z}y et al. 2011; Beck \& Wielebinski  2013; Drzazga et al. 2016). Our analysis indicates that radio deficiency in several WR galaxies is mainly due to lack of supernovae events. 

\subsection{Optical morphologies and tidal interaction features}

The optical morphologies of WR galaxies in our sample are studied using the SDSS and \Ha images. The color composite SDSS images reveal presence of blue star-forming disks in spiral galaxies. Most of the galaxies show very bright distinct blue regions. The fiber-based SDSS spectrum were taken mainly on these bright blue regions, which showed WR features. The \Ha emission corresponding to the blue regions is detected in all the galaxies. The \Ha and optical $r$-band morphologies of all galaxies in the sample were visually checked for signatures of tidal interactions. L\'opez-S\'anchez (2010) identified high occurrence of interaction related morphological features (plume, tails, merger and tidal dwarf galaxies), kinematical disturbances, and differences in chemical compositions of \HII regions in their sample of WR galaxies. The presence of multiple nuclei, arcs and tidal tails are generally considered as a signature of recent or ongoing tidal interaction (e.g., Beck \& Kovo 1999; L\'opez-S\'anchez et al. 2004a; Matsui et al. 2012; Adamo et al. 2012).  A summary of the prominent morphological features seen in WR galaxies in our sample is provided in Table~\ref{morphology}. The interaction probability in this table is inferred based on various morphological features. Galaxies showing \HI or optical tidal tails and distinct multiple nuclei with a disturbed optical envelope are termed as highly probable. Galaxies showing lopsidedness or asymmetric light distribution along with any other feature such as probable multiple nuclei or arcs are termed as moderately probable. Galaxies without any significant disturbed optical morphology are labeled with low interaction probability. It should be noted that several galaxies termed as having low probability show some interaction features such as bar, mild lopsidedness, or nuclear star formation. However, in absence of any strong tidal feature, we preferred to label these objects with low interaction probability.

In our sample, we find that 5 galaxies show prominent tidal tails or cometary shapes, 7 galaxies have bars, and 7 galaxies have two or three bright dominating \Ha regions suspected as multiple nuclei. The star formation dominated in the nuclear region is seen in 8 galaxies. In addition, two galaxies show presence of extra-planner \Ha regions. One galaxy has misaligned \Ha bar with respect to the optical bar.  A large number of 16 galaxies show lopsidedness (stellar light distribution or star formation) or irregular morphologies in our sample. Three galaxies show lopsided or disturbed radio continuum emission. Two galaxies are known to have disturbed \HI morphologies. It is not clear if multiple bright dominating \Ha regions are just different \HII regions or multiple nuclei of different galaxies such as other low mass dwarf galaxies in a merger stage. Kinematical and chemical abundance measurements will be useful to distinguish these two scenarios. The lopsidedness in galaxies has been studied previously in other galaxies, and is believed to be caused by tidal interactions (e.g., Jog 1997; Zaritsky \& Rix 1997; Angiras et al. 2006). The stellar bars can also result from recent tidal interaction events (e.g., Barnes \& Hernquist 1991; Miwa \& Noguchi 1998). The dwarf systems in our sample often show intense nuclear star-burst consistent with previous results (e.g., Strickland \& Stevens 1999; Adamo et al. 2011). It has been seen in N-body simulations (e.g., Hernquist \& Mihos 1995) that tidal interaction and mergers between galaxies can channel gas towards the center of galaxy as a result of loss of angular momentum. This gas can give rise to nuclear starburst in galaxies (Mihos \& Hernquist 1994). The optical morphological features seen in our study are also highly suggestive of prevalence of tidal interactions and mergers in the WR galaxies. A total of 16 galaxies are inferred here to be tidally interacting. These results are consistent with the results based on previous studies on WR galaxies, where tidal interactions were inferred in several WR galaxies (M\'endez \& Esteban 1999; L\'opez-S\'anchez \& Esteban 2008; Jaiswal \& Omar 2013; Karthick et al. 2014). All these results suggest that about two-third of WR galaxies show signatures of tidal interactions or mergers. It will be valuable to study environment of the WR galaxies, in terms of local density and velocity dispersion of galaxies in the surrounding region to probe tidal interactions. The GMRT (Giant Meterwave Radio Telescope) \HI 21cm-line observations have been completed in 20 galaxies from our sample. These \HI observations will help in understanding tidal interactions in these galaxies to greater depths.

\section{Conclusions}

Twenty-five Wolf-Rayet galaxies were imaged in the \Ha emission-line and $r$-band using 1.3-meter and 2-meter optical telescopes. These galaxies were selected from the SDSS with the distance cutoff at $\sim$25 Mpc. Majority of the selected galaxies in our sample are of small-size ($\le 10$ kpc) or dwarf galaxies.  The \Ha sensitivity obtained in our observations is typically $10^{-16} ~\rm erg~s^{-1}~cm^{-2}~arcsec^{-2}$. The integrated \Ha fluxes were corrected for internal and Galactic extinction, and also for the line contaminations in the \Ha and $r$ passbands. The star formation rates were derived for all the galaxies using H$\alpha$, FUV, FIR and radio luminosities. The archival data in the FUV, FIR and 1.4 GHz radio bands were taken from GALEX, IRAS and VLA surveys (FIRST and NVSS). We combined our sample with the sample of 20 WR galaxies studied by L\'opez-S\'anchez (2010). These two samples are complimentary to each other in the sense that our sample is dominated by galaxies with SFR $\le 1 M_\odot~{\rm yr}^{-1}$ while majority of galaxies in the L\'opez-S\'anchez (2010) sample has higher SFR values. 

The SFRs estimated from the luminosities in different wavebands were compared with the \Ha based SFR. The \Ha based SFR are found well correlated with other SFRs. We noticed, in general, that scatter in the correlations is higher towards the low SFR range. All the slopes in correlation with the \Ha SFR deviate from the unity. The highest deviation is seen with radio SFRs and the smallest deviation is with FUV based SFRs. These deviations and scatter are expected to be resulting mainly from uncertainties in the calibration constants for SFR-luminosity relations, star formation histories, age of the most recent star-burst, and assumptions about various physical processes in the ISM. These results are consistent, in general, with other similar studies. The radio-FIR correlation further reveals that majority of WR galaxies towards low SFR range are radio deficient, most likely due to lack of supernovae events in the young star-bursts in the WR galaxies. This result is also supported by N/O enrichment seen in several WR galaxies. 

The optical morphologies reveal presence of active star formation, wide-spread in the disks of spiral galaxies in our sample. All WR galaxies are dominated by a distinct region of very blue star forming region, coincident with the WR features detected in SDSS. Several small-size and dwarf galaxies in our sample show multiple \Ha emission region. It is not obvious if these multiple regions are just \HII regions or nuclei of separate dwarf galaxies in the merger state. Several galaxies in our sample show tidal features and lopsidedness in their stellar light distribution. Bars and nuclear star-bursts are also seen in a few galaxies. The morphologies of a total of 16 galaxies are inferred here to be suggestive of tidally interacting. In general, our study reveals tidal interactions as the prime reason for star formation trigger in WR galaxies, consistent with previous studies on WR galaxies (M\'endez \& Esteban 2000; L\'opez-S\'anchez \& Esteban 2008; L\'opez-S\'anchez 2010). Overall about two-thirds of WR galaxies in different samples show some signatures of tidal interactions or mergers. Further studies using the data presented in this paper are being carried out, in order to understand various SFR tracers in more details with an emphasis to effects of very young star-burst in WR galaxies. The \HI 21cm-line observations on 20 WR galaxies have been completed and will be used to understand tidal interaction process in greater details. Studies using long-slit optical spectroscopy are also planned to estimate chemical abundances in different \Ha emitting regions in order to reveal details of tidal interactions in these galaxies. 

\footnotesize
 
\section*{Acknowledgments} 

We thank the referee, A. R. L\'opez-S\'anchez for critically examining the manuscript, that greatly improved the clarity and contents of this paper. IRAF (Image Reduction and Analysis facility) is distributed by NOAO which is operated by AURA Inc., under cooperative agreement with NSF. This research has made use of the NASA/IPAC Extragalactic Database (NED) which is operated by the Jet Propulsion Laboratory, California Institute of Technology, under contract with the National Aeronautics and Space Administration. This research has made use of NASA's Astrophysics Data System. We thank the staff of ARIES, whose dedicated efforts made these observations possible. DFOT is run by Aryabhatta Research Institute of Observational Sciences with support from the Department of Science and Technology, Govt. of India. We wish to acknowledge the IUCAA/IGO staff for their support during our observations. We acknowledge the use of the SDSS. Funding for the SDSS and SDSS-II has been provided by the Alfred P. Sloan Foundation, the Participating Institutions, the National Science Foundation, the U.S. Department of Energy, the National Aeronautics and Space Administration, the Japanese Monbukagakusho, the Max Planck Society, and the Higher Education Funding Council for England. The SDSS Web Site is http://www.sdss.org/. The SDSS is managed by the Astrophysical Research Consortium for the Participating Institutions. The Participating Institutions are the American Museum of Natural History, Astrophysical Institute Potsdam, University of Basel, University of Cambridge, Case Western Reserve University, University of Chicago, Drexel University, Fermilab, the Institute for Advanced Study, the Japan Participation Group, Johns Hopkins University, the Joint Institute for Nuclear Astrophysics, the Kavli Institute for Particle Astrophysics and Cosmology, the Korean Scientist Group, the Chinese Academy of Sciences (LAMOST), Los Alamos National Laboratory, the Max-Planck-Institute for Astronomy (MPIA), the Max-Planck-Institute for Astrophysics (MPA), New Mexico State University, Ohio State University, University of Pittsburgh, University of Portsmouth, Princeton University, the United States Naval Observatory, and the University of Washington. GALEX (Galaxy Evolution Explorer) is a NASA Small Explorer, launched in 2003 April. We gratefully acknowledge NASA's support for construction, operation, and science analysis for the GALEX mission, developed in cooperation with the Centre National d'Etudes Spatiales of France and the Korean Ministry of Science and Technology. The Infrared Astronomical Satellite (IRAS) mission was a collaborative effort by the United States (NASA), the Netherlands (NIVR), and the United Kingdom (SERC). VLA (Very Large Array) is run by NRAO (National Radio Astronomy Observatory). The NRAO is a facility of the National Science Foundation operated under cooperative agreement by Associated Universities, Inc.

\section*{Appendix A: Star Formation Rate Calculations}

A brief overview of methods used to convert luminosities at different wavelengths to SFR is provided here.

\subsection*{A1: Far-ultraviolet emission} 

The ultraviolet (UV) radiation is emitted by the most massive stars with their surface temperatures  in the range of $T \sim 30000-60000$~K. The radiation shortward of the Lyman continuum ($\lambda \le 912$ \ang) is almost completely absorbed by the surrounding interstellar medium (ISM). However, the radiation longward of this limit ($\lambda > 912$ \ang) can escape the ISM and provides an estimate for the total UV radiation from the massive stars. The FUV (1344-1786~$\rm \mathring{A}$) band continuum emission can therefore be used to measure SFR in galaxies (e.g., Salim et al. 2007).

The FUV flux density is estimated here from the GALEX data. This flux density is corrected for foreground and internal extinctions. The GALEX FUV luminosity is converted to SFR using the relation given by Kennicutt (1998a):

\begin{equation}
SFR_{FUV} [M_\odot~{\rm yr}^{-1}] =\frac{L_{FUV} [{\rm erg~s^{-1} \mathring{A}^{-1}}]}{9.2\times 10^{39}}
\end{equation}

\subsection*{A2: \Ha emission} 

The absorbed FUV radiation shortward of the Lyman continuum limit ionizes the ISM, which in turn emits nebular emission lines. The strengths of nebular emission lines are directly proportional to the impinging FUV flux. The nebular lines therefore can provide an indirect measurement of the FUV flux from the most massive stars and hence in turn an estimate for the SFR. The \Ha line ($\lambda = 6563$ \ang), being the strongest among the nebular emission lines from \HII regions, is widely used to trace recent star formation in galaxies (e.g., Kennicutt 1998a). The SFR$_{H\alpha}$ is estimated here from the \Ha luminosity using the relation given Kennicutt (1998b) :

\begin{equation}
SFR_{H\alpha} [M_\odot~{\rm yr}^{-1}] = {{L_{H\alpha} [{\rm erg~s^{-1}}]} \over {1.26 \times 10^{41}}}
\end{equation}

\subsection*{A3: Far-infrared emission} 

The FUV radiation also heats the dust in the ISM. The dust emits thermal radiation in the far infrared bands ($40 - 120 ~\mu$m: Devereux \& Young 1990; Devereux \& Hameed 1997). The FIR emission therefore also indirectly traces the star formation rates in galaxies. The IRAS data provides the $60\mu$m and $100\mu$m flux densities, which can be converted into total FIR luminosity using the relations described in Yun et al. (2001):

\begin{equation}
{\rm log}~{L}_{60 \mu {\rm m}}[L_\odot] = 6.014 + 2~log~D + log~S_{60 \mu {\rm m}}
\end{equation}

\begin{equation}
L_{FIR}[L_\odot] = \left(1+{{S_{100 \mu {\rm m}}}\over{2.58~S_{60 \mu {\rm m}}}}\right) L_{60 \mu {\rm m}}[L_\odot]
\end{equation}

\noindent where $D$ is the distance in Mpc and $S_{60 \mu {\rm m}}$ and $S_{100\mu {\rm m}}$ are flux densities at $60\mu {\rm m}$ and $100\mu {\rm m}$, respectively in units of Jy. The SFR$_{FIR}$ can be then estimated using the relation given by Kennicutt (1998b) :

\begin{equation}
SFR_{FIR}[M_\odot~{\rm yr}^{-1}] = { {L_{FIR} [L_{\odot}]} \over {5.8 \times 10^9} }
\end{equation}

\subsection*{A4: 1.4~GHz radio emission} 

The most massive stars ($M \ge 8~M_\odot$) end their life in supernovae explosions within about 30 Myr of their formation. These powerful explosions accelerate cosmic electrons to relativistic speeds. These electrons in presence of galactic magnetic field emit synchrotron (non-thermal) radiation. This radiation follows a power law in frequency where spectral power at low frequencies is higher than that at higher frequencies. The NVSS and FIRST are the two very sensitive radio surveys carried out at 1.4~GHz using the VLA. Most of the understanding on radio emission from nearby galaxies is due to these radio surveys. The 1.4 GHz radio emission from normal star forming galaxies is dominated by the synchrotron emission. The free-free thermal emission from star forming regions also contributes at a level of $\sim$10\% to the total radio emission at 1.4 GHz. In absence of any contamination to the radio flux from active galactic nuclei (AGN), the radio flux can be used to estimate the rate of star formation in galaxies. A comprehensive review on the radio emission from galaxies can be found in Condon (1992). The radio synchrotron emission traces star formation over longer duration $\sim$100 Myr, i.e., typical lifetimes of relativistic cosmic electrons in galaxies. 

The 1.4~GHz radio continuum flux density obtained from NVSS and FIRST is converted into radio luminosity using the relation given by Yun et al. (2001):

\begin{equation}
{\rm log}~{L}_{1.4 \rm GHz}[{\rm W\,Hz^{-1}}] = 20.08 + 2~log~D + log~S_{1.4 \rm GHz}
\end{equation}

\noindent where $D$ is the distance in Mpc and $S_{1.4GHz}$ is the 1.4~GHz radio continuum flux density in Jy. The SFR$_{1.4 \rm GHz}$ is then estimated using the relation given by Condon et al. (2002):

\begin{equation}
SFR_{1.4 GHz}[M_\odot~{\rm yr}^{-1}] = \frac{L_{1.4 GHz}[{\rm W\,Hz^{-1}}]}{4.6\times 10^{21}}
\end{equation}

\onecolumn

\begin{landscape}
\begin{table*}
\centering
\caption{Basic properties of WR galaxies in our sample.}
\begin{tabular}{lccccccc}
\hline
\hline
Name & RA (J2000) & DEC (J2000)           & Type & $r$     & $v_{\mathrm{helio}}$ & Size     & Other Names \\
     & h m s      & $^{\circ}$ $'$ $''$   &      & [mag]   & [km/s]               & [arcmin] &             \\
\hline 
\noalign{\smallskip}
UM~311        & 01 15 34.4 & $-$00 51 46 & BCD             & $17.83$     & $1675 \pm 2$  & 0.11 & SHOC~056 \\
IC~225        & 02 26 28.3 & $+$01 09 38 & E               & $14.05$     & $1535 \pm 2$  & 0.94 & UGC~1907, MRK~1038 \\
NGC~941       & 02 28 27.8 & $-$01 09 06 & SAB(rs)c        & $12.91$     & $1608 \pm 6^a$& 1.83 & UGC~1954 \\
NGC~1087      & 02 46 25.1 & $-$00 29 55 & SAB(rs)c        & $11.69$     & $1517 \pm 4^b$& 3.12 & UGC~2245 \\
UGCA~116      & 05 55 42.6 & $+$03 23 32 & compact Irr pec & $11.25^A$   & $789 \pm 4^c$ & 0.56 & II~Zw~40 \\
UGCA~130      & 06 42 15.5 & $+$75 37 33 & Irr             & $14.69^A$   & $792 \pm 5^c$ & 0.70 & MRK~5 \\
NGC~2799      & 09 17 31.0 & $+$41 59 39 & SB(s)m          & $18.63$     & $1673 \pm 4^d$& 1.90 & UGC~4909 \\
MRK~22        & 09 49 30.3 & $+$55 34 47 & BCD             & $15.67$     & $1551 \pm 12^e$& 0.48& UGCA~184 \\
IC~2524       & 09 57 32.8 & $+$33 37 11 & S               & $14.33$     & $1450 \pm 3$  & 0.74 & MRK~411 \\
KUG~1013+381  & 10 16 24.5 & $+$37 54 46 & BCD             & $16.00$     & $1173 \pm 3$  & 0.38 & -- \\
NGC~3294      & 10 36 16.2 & $+$37 19 29 & SA(s)c          & $11.54$     & $1586 \pm 6^c$& 2.30 & UGC~5753 \\
NGC~3381      & 10 48 24.8 & $+$34 42 41 & SB pec          & $12.84$     & $1629 \pm 2^f$& 1.46 & UGC~5909 \\
NGC~3423      & 10 51 14.3 & $+$05 50 24 & SA(s)cd         & $11.93$     & $1011 \pm 5^a$& 3.80 & UGC~5962 \\
NGC~3430      & 10 52 11.4 & $+$32 57 02 & SAB(rs)c        & $12.14$     & $1586 \pm 1^g$& 4.00 & UGC~5982 \\
CGCG~038-051  & 10 55 39.2 & $+$02 23 45 & dIrr            & $16.04$     & $1021 \pm 2$  & 0.57 & -- \\
IC~2828       & 11 27 10.9 & $+$08 43 52 & Im              & $14.71$     & $1039 \pm 12^h$& 0.73& CGCG~067-078 \\
UM~439        & 11 36 36.8 & $+$00 48 58 & Irr             & $14.23^A$   & $1099 \pm 4^i$& 0.72 & UGC~6578 \\
NGC~3949      & 11 53 41.7 & $+$47 51 31 & SA(s)bc         & $11.22$     & $800 \pm 1^j$ & 2.92 & UGC~6869 \\
I~SZ~59       & 11 57 28.0 & $-$19 37 27 & S0              & $14.24^A$   & $2135 \pm 18^k$& 1.40& ESO~572-~G ~025 \\
CGCG~041-023  & 12 01 44.3 & $+$05 49 17 & SB              & $14.73$     & $1350 \pm 2^l$& 0.72 & VV~462 \\
SBS~1222+614  & 12 25 05.4 & $+$61 09 11 & dIrr            & $14.70$     & $706 \pm 2^e$ & 0.51 & -- \\
NGC~4389      & 12 25 35.1 & $+$45 41 05 & SB(rs)bc pec    & $12.07$     & $718 \pm 1^j$ & 2.14 & UGC~7514 \\
IC~3521       & 12 34 39.5 & $+$07 09 37 & IBm             & $13.24$     & $595 \pm 6^m$ & 1.43 & UGC~7736 \\
UGC~9273      & 14 28 10.8 & $+$13 33 06 & Im              & $15.03$     & $1289 \pm 5^n$& 0.97 & CGCG~075-042 \\
MRK~475       & 14 39 05.4 & $+$36 48 22 & BCD             & $16.30$     & $583 \pm 2^e$ & 0.36 & -- \\
\noalign{\smallskip}    
\hline
\end{tabular}
\begin{flushleft}
{\footnotesize {References for the $v_{\mathrm{helio}}$: $^a$ HI Parkes All Sky Survey (HIPASS: Barnes et al. 2001), $^b$ Koribalski et al. (2004), $^c$ Third Reference Catalogue (RC3: de Vaucouleurs et al. 1991), $^d$ Monnier Ragaigne et al. (2003), $^e$ Thuan et al. (1999), $^f$ van Driel et al. (2001), $^g$ Nordgren et al. (1997), $^h$ Smoker et al. (2000), $^i$ Comte et al. (1999), $^j$ Verheijen \& Sancisi (2001), $^k$ Firth et al. (2006), $^l$ Lu et al. (1993), $^m$ Binggeli et al. (1993), $^n$ Schneider et al. (1990). The $v_{\mathrm{helio}}$ values for other galaxies are taken from SDSS catalogues (Abazajian et al. 2009).\linebreak}}\\

{\footnotesize {The $r$-band magnitudes super-scripted by $A$ are calculated using Johnson $B$ and Cousins $R$ band magnitudes from Gil de Paz et al. (2003), and Lupton transformation equations. The magnitudes for other galaxies are taken from the SDSS (Abazajian et al. 2009).}}
\end{flushleft}
\label{sample}
\end{table*}
\end{landscape}

\begin{table*}
\centering
\caption{Extinction parameters and metallicities.}
\footnotesize
\begin{tabular}{lcccc}
\hline
\hline
Galaxy name & $f_{H\alpha}/f_{H\beta}$ & $E_g(B-V)$ & $E_f(B-V)$ & 12 + log(O/H) \\
     &                          & [mag]      & [mag]      &      \\
\hline 
\noalign{\smallskip}
UM~311       & $3.58 \pm 0.09$ & $0.23 \pm 0.03$ & $0.03$ & $8.31 \pm 0.04^f$ \\
IC~225       & $3.59 \pm 0.06$ & $0.23 \pm 0.02$ & $0.03$ & $8.52 \pm 0.02$ \\
NGC~941      & $3.21 \pm 0.10$ & $0.12 \pm 0.03$ & $0.03$ & $8.68 \pm 0.37$ \\
NGC~1087     & $4.26 \pm 0.15$ & $0.40 \pm 0.04$ & $0.03$ & $8.80 \pm 0.04$ \\
UGCA~116     & $6.98 \pm 0.05^a$ &       ---     & $0.72$ & $8.09 \pm 0.02^g$ \\
UGCA~130     & $4.08 \pm 0.05^b$ &       ---     & $0.07$ & $8.04 \pm 0.04^h$ \\
NGC~2799     & $3.98 \pm 0.11$ & $0.33 \pm 0.03$ & $0.02$ & $8.74 \pm 0.02$ \\
MRK~22       & $3.03 \pm 0.13$ & $0.06 \pm 0.04$ & $0.01$ & $8.04 \pm 0.01^i$ \\
IC~2524      & $3.27 \pm 0.08$ & $0.14 \pm 0.02$ & $0.01$ & $8.39 \pm 0.03$ \\
KUG~1013+381 & $3.70 \pm 0.04^c$ &       ---     & $0.01$ & $7.50 \pm 0.01$ \\
NGC~3294     & $4.11 \pm 0.40$ & $0.37 \pm 0.10$ & $0.02$ & $8.79 \pm 0.01$ \\
NGC~3381     & $3.38 \pm 0.09$ & $0.17 \pm 0.03$ & $0.02$ & $8.84 \pm 0.01$ \\
NGC~3423     & $3.16 \pm 0.37$ & $0.10 \pm 0.12$ & $0.03$ & $8.41 \pm 0.19$ \\
NGC~3430     & $4.02 \pm 0.16$ & $0.34 \pm 0.04$ & $0.02$ & $8.77 \pm 0.01$ \\
CGCG~038-051 & $3.20 \pm 0.10$ & $0.11 \pm 0.03$ & $0.03$ & $7.91 \pm 0.08$ \\
IC~2828      & $3.44 \pm 0.15$ & $0.19 \pm 0.04$ & $0.05$ & $8.33 \pm 0.01$ \\
UM~439       & $3.22 \pm 0.08$ & $0.12 \pm 0.03$ & $0.02$ & $8.08 \pm 0.03^j$ \\
NGC~3949     & $3.77 \pm 0.09$ & $0.28 \pm 0.02$ & $0.02$ & $8.46 \pm 0.02$ \\
I~SZ~59      & $4.06 \pm 0.48^d$ &       ---     & $0.04$ & $\sim 8.40^k$   \\
CGCG~041-023 & $3.44 \pm 0.31$ & $0.19 \pm 0.09$ & $0.01$ & $8.25 \pm 0.01$ \\
SBS~1222+614 & $2.90 \pm 0.08$ & $0.01 \pm 0.03$ & $0.01$ & $8.00 \pm 0.02^l$ \\
NGC~4389     & $3.80 \pm 0.10$ & $0.29 \pm 0.03$ & $0.01$ & $8.85 \pm 0.01$ \\
IC~3521      & $3.74 \pm 0.42$ & $0.27 \pm 0.11$ & $0.02$ & $8.88 \pm 0.03$ \\
UGC~9273     & $3.01 \pm 0.08$ & $0.05 \pm 0.03$ & $0.02$ & $8.33 \pm 0.01$ \\
MRK~475      & $3.15 \pm 0.03^e$ &       ---     & $0.01$ & $7.97 \pm 0.02^m$ \\
\noalign{\smallskip}    
\hline
\end{tabular}
\begin{flushleft}
{\footnotesize {$^a$ $f_{H\alpha}/f_{H\beta} = 6.975 \pm 0.052$, $E(B-V) = 0.90 \pm 0.01$ (Guseva, Izotov \& Thuan 2000).\newline
$^b$ $f_{H\alpha}/f_{H\beta} = 4.080 \pm 0.046$, $E(B-V) = 0.36 \pm 0.01$ (Izotov \& Thuan 1998).\newline
$^c$ $f_{H\alpha}/f_{H\beta} = 3.703 \pm 0.043$, $E(B-V) = 0.26 \pm 0.01$ (from our spectroscopic measurements).\newline 
$^d$ $f_{H\alpha}/f_{H\beta} = 4.061 \pm 0.484$, $E(B-V) = 0.35 \pm 0.12$ (from our spectroscopic measurements).\newline
$^e$ $f_{H\alpha}/f_{H\beta} = 3.148 \pm 0.031$, $E(B-V) = 0.10 \pm 0.01$ (Izotov, Thuan \& Lipovetsky 1994).\linebreak}}\\

{\footnotesize {References for 12 + log(O/H): \quad $^f$ Izotov \& Thuan (1998), $^g$ Guseva, Izotov \& Thuan (2000), $^h$ Izotov \& Thuan (1998), $^i$ Izotov, Thuan \& Lipovetsky (1994), $^j$ Zhao et al. (2013), $^k$ Kunth \& Joubert (1985), $^l$ Ekta \& Chengalur (2010), $^m$ Izotov, Thuan \& Lipovetsky (1994). The oxygen abundances for other galaxies are taken from Brinchmann et al. (2008).}}
\end{flushleft}
\label{archive}
\end{table*}

\begin{landscape}
\begin{table*}
\centering
\caption{Summary of the optical observations.}
\begin{tabular}{lcccccccc}
\hline
\hline
Galaxy name & Telescope & Date & Exposure in \Ha & Exposure in $r$ & FWHM PSF & \Ha Sensitivity                          & $\lambda_{\mathrm{H}\alpha}$ & $\lambda_0$\\
     &           &      & [min]             & [min]             & [arcsec] & $10^{-16}\left[\frac{\mathrm{erg}~\mathrm{s}^{-1}}{\mathrm{cm}^{2}~\mathrm{arcsec}^{2}}\right]$ & [\ang] & [\ang]\\
\hline 
UM~311       & DFOT & 2012 Nov 09    & 150 & 10 & 2.0 & 1.47 & 6599.7 & 6570\\
IC~225       & DFOT & 2012 Dec 08,09 & 220 & 30 & 2.3 & 0.88 & 6596.6 & 6570\\
NGC~941      & DFOT & 2012 Dec 07,08 & 195 & 38 & 2.2 & 0.42 & 6598.2 & 6570\\
NGC~1087     & DFOT & 2012 Nov 08,09 & 170 & 22 & 1.9 & 1.15 & 6596.2 & 6570\\
UGCA~116     & DFOT & 2012 Nov 08,09 & 215 & 32 & 2.2 & 1.15 & 6580.3 & 6570\\
UGCA~130     & DFOT & 2012 Dec 07    & 165 & 35 & 2.6 & 1.21 & 6580.3 & 6570\\
NGC~2799     & DFOT & 2013 Feb 07    & 155 & 25 & 2.1 & 0.85 & 6599.6 & 6563\\
MRK~22       & DFOT & 2012 Dec 08,09 & 235 & 50 & 2.1 & 0.92 & 6597.0 & 6570\\
IC~2524      & IGO  & 2012 Mar 19    & 117 & 15 & 1.3 & 1.20 & 6594.7 & 6563\\
KUG~1013+381 & IGO  & 2012 Mar 20    & 150 & 20 & 1.4 & 1.32 & 6588.7 & 6563\\
NGC~3294     & DFOT & 2013 Feb 08    & 155 & 15 & 2.2 & 0.85 & 6597.7 & 6563\\
NGC~3381     & DFOT & 2012 Apr 23    & 110 & 15 & 2.1 & 1.80 & 6598.7 & 6570\\
NGC~3423     & DFOT & 2013 Mar 12    & 150 & 20 & 1.9 & 0.96 & 6585.1 & 6563\\
NGC~3430     & DFOT & 2013 Mar 11    & 140 & 30 & 2.0 & 0.86 & 6597.7 & 6563\\
CGCG~038-051 & DFOT & 2013 Mar 07    & 155 & 25 & 2.2 & 0.85 & 6585.4 & 6563\\
IC~2828      & IGO  & 2012 Mar 20    & 130 & 20 & 1.2 & 1.13 & 6585.7 & 6563\\
UM~439       & DFOT & 2012 Apr 22,23 & 105 & 20 & 2.1 & 1.80 & 6587.1 & 6570\\
NGC~3949     & DFOT & 2013 Mar 08    & 175 & 30 & 1.9 & 0.78 & 6580.5 & 6563\\
I~SZ~59      & DFOT & 2013 Mar 10    & 150 & 25 & 2.2 & 1.03 & 6609.7 & 6563\\
CGCG~041-023 & DFOT & 2013 Feb 08    & 165 & 15 & 2.2 & 0.89 & 6592.6 & 6563\\
SBS~1222+614 & DFOT & 2013 Mar 08,10 & 150 & 25 & 2.1 & 0.79 & 6578.5 & 6563\\
NGC~4389     & DFOT & 2013 Mar 07,08 & 160 & 30 & 1.9 & 0.74 & 6578.7 & 6563\\
IC~3521      & DFOT & 2012 Apr 22    & 89  & 15 & 2.5 & 1.90 & 6576.0 & 6570\\
UGC~9273     & IGO  & 2012 Mar 19    & 140 & 20 & 1.2 & 1.03 & 6591.2 & 6563\\
MRK~475      & DFOT & 2012 May 20,21 & 170 & 23 & 2.2 & 0.98 & 6575.8 & 6570\\
\hline
\end{tabular}
\label{obs}
\end{table*}
\end{landscape}

\begin{table*}
\centering
\caption{Instrument calibration results.}
\begin{tabular}{cccccc}
\hline
\hline
Telescope & $\lambda_0$ for \Ha filter & FWHM for \Ha filter & $C_{H\alpha}$ & $k_{H\alpha}$ & WNCR\\
          & [\ang]                     & [\ang]              & 10$^{-16} \left[\rm \frac{erg~s^{-1}cm^{-2}}{counts~s^{-1}}\right]$ & [mag/airmass] &\\
\hline 
DFOT & 6570 & 77  & $31.3 \pm 0.5$  & $0.12 \pm 0.03$ & $17.2 \pm 0.8$\\
DFOT & 6563 & 100 & $35.3 \pm 0.6$  & $0.12 \pm 0.02$ & $13.8 \pm 0.6$\\
IGO  & 6563 & 80  & $7.3 \pm 0.3$   & $0.30 \pm 0.03$ & $14.6 \pm 0.6$\\
\hline
\end{tabular}
\label{cal}
\end{table*}

\begin{table*}
\centering
\caption{FUV, FIR and radio flux.}
\begin{tabular}{lcccc}
\hline
\hline
Galaxy name & $F_{FUV}$ & $S_{60 \mu {\rm m}}$ & $S_{100 \mu {\rm m}}$ & $S_{1.4 \rm GHz}$ \\
     & [$10^{-14} {\rm erg~s^{-1}~cm^{-2}~\mathring{A}^{-1}}$] & [Jy] & [Jy] & [mJy] \\
\hline 
\noalign{\smallskip}
UM~311       &	$1.4 \pm 0.1$    &   ---              & ---                 & $0.6 \pm 0.1^A$    \\
IC~225       &  $1.3 \pm 0.1$    &   ---              & ---                 & $<0.4^A$           \\
NGC~941      &	$9.1 \pm 0.9$    &   $0.9 \pm 0.1^a$  & $2.5 \pm 0.2^a$     & $7.6 \pm 2.2^e$    \\
NGC~1087     &	$64.0 \pm 6.7$   &   $12.2 \pm 0.1^b$ & $27.3 \pm 0.2^b$    & $\sim 136.0^f$     \\ 
UGCA~116     &	$3.4 \pm 0.3$    &   $6.0 \pm 0.4^a$  & $5.3 \pm  0.8^c$    & $34.2 \pm 1.4^e$   \\
UGCA~130     &	$1.7 \pm 0.2$    &   $0.21 \pm 0.04^a$& $<0.8^a$            & $<1.0^g$           \\
NGC~2799     &	$4.0 \pm 0.4$    &   $<1.5^d$         & $<1.0^d$            & $\sim 4.1^f$       \\
MRK~22       &	$0.7 \pm 0.1$    &   ---              & ---                 & $1.4 \pm 0.3^A$    \\
IC~2524      &	$1.6 \pm 0.2$    &   $0.30 \pm 0.04^a$& $0.5 \pm 0.1^a$     & $0.7 \pm 0.2^A$    \\
KUG~1013+381 &	$2.6 \pm 0.3$    &   ---              & ---                 & $1.3 \pm 0.3^A$    \\
NGC~3294     &	$22.6 \pm 2.3$   &   $6.2 \pm 0.3^a$  & $16.8 \pm 0.1^a$    & $53.8 \pm 2.6^e$   \\ 
NGC~3381     &	$8.7 \pm 0.9$    &   $1.4 \pm 0.1^a$  & $3.9 \pm 0.2^a$     & $6.2 \pm 0.6^e$    \\
NGC~3423     &	$29.4 \pm 3.0$   &   ---              & ---                 & $\sim 24.9^f$      \\ 
NGC~3430     &	$30.0 \pm 3.2$   &   $3.1 \pm 0.2^a$  & $9.2 \pm 0.8^a$     & $35.3 \pm 2.5^e$   \\ 
CGCG~038-051 &	$0.8 \pm 0.1$    &   ---              & ---                 & $14.7 \pm 0.5^A$   \\
IC~2828      &	$2.1 \pm 0.2$    &   $0.3 \pm 0.1^a$  & $0.7 \pm 0.2^a$     & $<0.4^A$           \\
UM~439       &	$3.3 \pm 0.3$    &   $0.36 \pm 0.05^a$& $<0.8^a$            & $4.7 \pm 0.5^A$    \\
NGC~3949     &	$43.2 \pm 4.5$   &   $11.37 \pm 0.04^a$ & $24.9 \pm 1.2^a$  & $120.6 \pm 4.4^e$  \\ 
I~SZ~59      &	    ---          &   $0.5 \pm 0.1^a$  & $0.9 \pm 0.2^a$     & ---                \\
CGCG~041-023 &      ---          &   ---              & ---                 & $4.4 \pm 0.3^A$    \\
SBS~1222+614 &	$1.8 \pm 0.2$    &   $0.29 \pm 0.04^a$& $<0.5^a$            & $<0.5^A$           \\
NGC~4389     &	$11.8 \pm 1.2$   &   ---              & ---                 & $20.8 \pm 1.7^e$   \\ 
IC~3521      &	$2.6 \pm 0.3$    &   $1.0 \pm 0.1^a$  & $2.3 \pm 0.3^a$     & $3.7 \pm 0.3^A$    \\
UGC~9273     &	$1.3 \pm 0.3$    &   ---              & ---                 & $<0.4^A$           \\
MRK~475      &	$0.6 \pm 0.1$    &   ---              & ---                 & $<0.4^A$           \\
\noalign{\smallskip}    
\hline
\end{tabular}
\begin{flushleft}
{\footnotesize {References for $S_{60 \mu {\rm m}}$ and $S_{100 \mu {\rm m}}$: $^a$ Moshir et al. (1990), $^b$ Soifer (1989), $^c$ Sanders et al. (2003), $^d$~Surace et al. (2004)\linebreak}}\\

{\footnotesize {References for $S_{1.4 \rm GHz}$: $^e$ Condon et al. (1998), $^f$ Condon et al. (2002), $^g$ Leroy et al. (2005). The values denoted by superscript `$A$' are estimated using the FIRST images, while the others are taken from the NVSS catalogue.}}
\end{flushleft}
\label{flux_a}
\end{table*}

\begin{landscape}
\begin{table*}
\centering
\caption{The \Ha flux.}
\begin{tabular}{lcccc}
\hline
\hline
Galaxy name & $F_{H\!\alpha}$ & $F^i_{H\!\alpha}$ & $F^{ii}_{H\!\alpha}$ & $F^{iii}_{H\!\alpha}$ \\
     & [10$^{-14}$erg~s$^{-1}$cm$^{-2}$] & [10$^{-14}$erg~s$^{-1}$cm$^{-2}$]               & [10$^{-14}$erg~s$^{-1}$cm$^{-2}$] & [10$^{-14}$erg~s$^{-1}$cm$^{-2}$] \\
\hline
UM~311       & $7.9 \pm 0.5$    & $7.6 \pm 0.5$    & $8.5 \pm 0.5$   & $15.8 \pm 1.1$    \\
IC~225       & $9.2 \pm 3.4$    & $7.8 \pm 3.4$    & $8.9 \pm 3.4$   & $16.2 \pm 6.2$    \\
NGC~941      & $99.0 \pm 6.4$   & $85.5 \pm 6.5$   & $95.9 \pm 6.6$  & $136 \pm 12$      \\
NGC~1087     & $300 \pm 40$     & $240 \pm 41$     & $274 \pm 41$    & $749 \pm 115$     \\
UGCA~116     & $191 \pm 11$     & $188 \pm 11$     & $200 \pm 11$    & $1645 \pm 145$    \\
UGCA~130     & $19.2 \pm 2.4$   & $18.7 \pm 2.4$   & $20.1 \pm 2.4$  & $46.7 \pm 5.6$    \\
NGC~2799     & $6.7 \pm 1.1$    & $5.7 \pm 1.1$    & $6.5 \pm 1.1$   & $14.6 \pm 2.5$    \\
MRK~22       & $8.5 \pm 0.8$    & $8.3 \pm 0.8$    & $9.1 \pm 0.8$   & $10.7 \pm 1.3$    \\
IC~2524      & $12.6 \pm 3.2$   & $12.1 \pm 3.2$   & $12.8 \pm 3.2$  & $18.2 \pm 4.6$    \\
KUG~1013+381 & $25.4 \pm 1.8$   & $25.3 \pm 1.8$   & $26.4 \pm 1.8$  & $48.5 \pm 3.4$    \\
NGC~3294     & $283 \pm 60$     & $231 \pm 60$     & $262 \pm 60$    & $648 \pm 161$     \\
NGC~3381     & $31.4 \pm 12.6$  & $25.6 \pm 12.6$  & $29.5 \pm 12.6$ & $45.7 \pm 19.6$   \\
NGC~3423     & $354 \pm 51$     & $306 \pm 52$     & $338 \pm 52$    & $454 \pm 122$     \\
NGC~3430     & $241 \pm 37$     & $197 \pm 37$     & $224 \pm 37$    & $519 \pm 89$      \\
CGCG~038-051 & $19.5 \pm 1.5$   & $19.0 \pm 1.5$   & $20.6 \pm 1.5$  & $28.7 \pm 2.6$    \\
IC~2828      & $37.7 \pm 3.4$   & $35.5 \pm 3.4$   & $37.3 \pm 3.4$  & $64.7 \pm 7.1$    \\
UM~439       & $27.3 \pm 2.5$   & $26.7 \pm 2.5$   & $28.8 \pm 2.5$  & $40.0 \pm 4.1$    \\
NGC~3949     & $509 \pm 61$     & $419 \pm 62$     & $464 \pm 62$    & $932 \pm 126$     \\
I~SZ~59      & $48.1 \pm 3.4$   & $47.6 \pm 3.4$   & $53.8 \pm 3.5$  & $122 \pm 17$      \\
CGCG~041-023 & $28.7 \pm 2.8$   & $28.0 \pm 2.8$   & $30.5 \pm 2.8$  & $48.9 \pm 8.0$    \\
SBS~1222+614 & $89.1 \pm 4.7$   & $87.7 \pm 4.7$   & $94.6 \pm 4.8$  & $100 \pm 9$       \\
NGC~4389     & $170 \pm 28$     & $130 \pm 28$     & $144 \pm 28$    & $293 \pm 58$      \\
IC~3521      & $25.8 \pm 5.8$   & $20.2 \pm 5.8$   & $22.0 \pm 5.8$  & $43.2 \pm 12.8$   \\
UGC~9273     & $12.8 \pm 2.1$   & $12.3 \pm 2.1$   & $12.9 \pm 2.1$  & $15.1 \pm 2.6$    \\
MRK~475      & $32.1 \pm 1.8$   & $31.7 \pm 1.8$   & $33.8 \pm 1.8$  & $42.7 \pm 2.4$    \\
\hline
\end{tabular}
\label{flux}
\end{table*}
\end{landscape}

\begin{landscape}
\begin{table*}
\centering
\caption{The luminosities and SFRs in different wave-bands.}
\footnotesize
\begin{tabular}{lccccccccc}
\hline
\hline
Galaxy name & $L_{FUV}$ & $L_{H\!\alpha}$ & $L_{FIR}$ & $L_{1.4GHz}$ & $SFR_{FUV}$ & $SFR_{H\!\alpha}$ & $SFR_{FIR}$ & $SFR_{1.4GHz}$ & $SFR_{assumed}$\\
     & ${\rm [10^{39}erg~s^{-1} \mathring{A}^{-1}]}$ & ${\rm [10^{40}erg~s^{-1}]}$ & $[10^{8}L_\odot]$ & $[10^{20}{\rm W\,Hz^{-1}}]$ & $[M_\odot~{\rm yr}^{-1}]$ & $[M_\odot~{\rm yr}^{-1}]$ & $[M_\odot~{\rm yr}^{-1}]$ & $[M_\odot~{\rm yr}^{-1}]$ & $[M_\odot~{\rm yr}^{-1}]$ \\
\hline 
UM~311       & $0.8 \pm 0.1$  & $0.9 \pm 0.1$   & --            & $0.4 \pm 0.1$  & $0.09 \pm 0.01$  & $0.07 \pm 0.01$ & --                & $0.009 \pm 0.002$  & $0.07$\\
IC~225       & $0.7 \pm 0.1$  & $0.8 \pm 0.4$   & --            & $<$0.20        & $0.08 \pm 0.01$  & $0.06 \pm 0.03$ & --                & $<$0.004           & $0.06$\\
NGC~941      & $5.0 \pm 0.8$  & $7.5 \pm 1.1$   & $8.8 \pm 1.5$ & $4.20 \pm 1.4$ & $0.54 \pm 0.09$  & $0.60 \pm 0.08$ & $0.15 \pm 0.03$   & $0.09 \pm 0.03$    & $0.35$\\
NGC~1087     & $31.7 \pm 5.0$ & $36.7 \pm 7.6$  & $96.4 \pm 6.0$& $66.9 \pm 3.6$ & $3.45 \pm 0.54$  & $2.91 \pm 0.60$ & $1.66 \pm 0.10$   & $1.45 \pm 0.08$    & $2.29$\\
UGCA~116     & $0.5 \pm 0.1$  & $21.8 \pm 3.1$  & $9.2 \pm 1.1$ & $4.6 \pm 0.4$  & $0.05 \pm 0.01$  & $1.73 \pm 0.25$ & $0.16 \pm 0.02$   & $0.10 \pm 0.01$    & $0.13$\\
UGCA~130     & $0.25 \pm 0.04$& $0.6 \pm 0.1$   & $<$0.6        & $<$0.13        & $0.018 \pm 0.002$& $0.05 \pm 0.01$ & $<$0.01            & $<$0.003         & $0.01$\\
NGC~2799     & $2.4 \pm 0.4$  & $0.9 \pm 0.2$   & $<$9.7        & $2.5 \pm 0.1$  & $0.26 \pm 0.04$  & $0.07 \pm 0.02$ & $<$0.17           & $0.054 \pm 0.002$  & $0.12$\\
MRK~22       & $0.4 \pm 0.1$  & $0.5 \pm 0.1$   & --            & $0.7 \pm 0.2$  & $0.04 \pm 0.01$  & $0.04 \pm 0.01$ & --                & $0.015 \pm 0.004$  & $0.04$\\
IC~2524      & $0.7 \pm 0.1$  & $0.8 \pm 0.3$   & $2.0 \pm 0.4$ & $0.3 \pm 0.1$  & $0.08 \pm 0.01$  & $0.06 \pm 0.02$ & $0.03 \pm 0.01$   & $0.007 \pm 0.002$  & $0.05$\\
KUG~1013+381 & $0.8 \pm 0.1$  & $1.4 \pm 0.2$   & --            & $0.4 \pm 0.1$  & $0.08 \pm 0.01$  & $0.11 \pm 0.01$ & --                & $0.009 \pm 0.002$  & $0.08$\\
NGC~3294     & $12.1 \pm 1.9$ & $34.4 \pm 10.5$ & $58.7 \pm 6.0$& $28.9 \pm 3.0$ & $1.32 \pm 0.21$  & $2.73 \pm 0.83$ & $1.01 \pm 0.10$   & $0.63 \pm 0.07$    & $1.17$\\
NGC~3381     & $4.9 \pm 0.8$  & $2.6 \pm 1.2$   & $14.1 \pm 1.8$& $3.5 \pm 0.5$  & $0.53 \pm 0.09$  & $0.21 \pm 0.10$ & $0.24 \pm 0.03$   & $0.08 \pm 0.01$    & $0.23$\\
NGC~3423     & $6.4 \pm 1.0$  & $9.9 \pm 3.2$   & --            & $5.4 \pm 0.3$  & $0.70 \pm 0.11$  & $0.79 \pm 0.25$ & --                & $0.12 \pm 0.01$    & $0.70$\\
NGC~3430     & $16.3 \pm 2.6$ & $27.8 \pm 6.2$  & $30.7 \pm 3.6$& $19.0 \pm 2.4$ & $1.77 \pm 0.28$  & $2.21 \pm 0.49$ & $0.53 \pm 0.06$   & $0.41 \pm 0.05$    & $1.15$\\
CGCG~038-051 & $0.18 \pm 0.03$& $0.6 \pm 0.1$   & --            & $3.3 \pm 0.3$  & $0.020 \pm 0.003$& $0.05 \pm 0.01$ & --                 & $0.07 \pm 0.01$  & $0.05$\\
IC~2828      & $0.5 \pm 0.1$  & $1.5 \pm 0.3$   & $1.1 \pm 0.4$ & $<$0.09        & $0.05 \pm 0.01$  & $0.12 \pm 0.02$ & $0.02 \pm 0.01$   & $<$0.002           & $0.04$\\
UM~439       & $0.9 \pm 0.1$  & $1.0 \pm 0.2$   & $<$1.5        & $1.2 \pm 0.2$  & $0.10 \pm 0.01$  & $0.08 \pm 0.01$ & $<$0.03           & $0.026 \pm 0.004$  & $0.06$\\
NGC~3949     & $5.9 \pm 0.9$  & $12.7 \pm 2.4$  & $24.7 \pm 1.4$& $16.5 \pm 1.5$ & $0.64 \pm 0.10$  & $1.01 \pm 0.19$ & $0.43 \pm 0.02$   & $0.36 \pm 0.03$    & $0.54$\\
I~SZ~59      & --             & $11.8 \pm 2.3$  & $7.1 \pm 1.8$ & --             & --               & $0.94 \pm 0.18$ & $0.12 \pm 0.03$   & --                 & $0.53$\\
CGCG~041-023 & --             & $1.9 \pm 0.4$   & --              & $1.7 \pm 0.2$  & --               & $0.15 \pm 0.03$ & --                & $0.037 \pm 0.004$  & $0.09$\\
SBS~1222+614 & $0.19 \pm 0.03$& $1.1 \pm 0.2$   & $<$0.5        & $<$0.05        & $0.021 \pm 0.003$& $0.09 \pm 0.01$ & $<$0.01           & $<$0.001           & $0.02$\\
NGC~4389     & $1.3 \pm 0.2$  & $3.2 \pm 0.8$   & --            & $2.3 \pm 0.3$  & $0.14 \pm 0.02$  & $0.25 \pm 0.06$ & --                & $0.05 \pm 0.01$  & $0.14$\\
IC~3521      & $0.20 \pm 0.03$& $0.3 \pm 0.1$   & $1.3 \pm 0.2$ & $0.28 \pm 0.04$& $0.022 \pm 0.003$& $0.03 \pm 0.01$ & $0.022 \pm 0.003$ & $0.006 \pm 0.001$ & $0.02$\\
UGC~9273     & $0.5 \pm 0.1$  & $0.5 \pm 0.1$   & --            & $<$0.12        & $0.05 \pm 0.01$  & $0.04 \pm 0.01$ & --                & $<$0.003          & $0.04$\\
MRK~475      & $0.05 \pm 0.01$& $0.31 \pm 0.03$ & --            & $<$0.03        & $0.005 \pm 0.001$& $0.025 \pm 0.003$& --               & $<$0.001          & $0.01$\\
\hline
\end{tabular}
\label{sfr}
\end{table*}
\end{landscape}

\begin{table*}
\centering
\caption{The $q$ parameter and non-thermal to thermal radio flux ratio R.}
\begin{tabular}{lcc}
\hline
\hline
Name & $q$ & log $R$\\
\hline 
UM~311       & --              & $0.33 \pm 0.11$ \\
IC~225       & --              & --              \\
NGC~941      & $2.33 \pm 0.30$ & $0.56 \pm 0.17$ \\
NGC~1087     & $2.16 \pm 0.10$ & $1.15 \pm 0.08$ \\
UGCA~116     & $2.31 \pm 0.07$ & $-0.14 \pm 0.08$\\
UGCA~130     & $\sim2.65$      & --              \\
NGC~2799     & $< 2.60$        & $1.35 \pm 0.09$ \\
MRK~22       & --              & $0.99 \pm 0.12$ \\
IC~2524      & $2.79 \pm 0.31$ & $0.34 \pm 0.22$ \\
KUG~1013+381 & --              & $0.09 \pm 0.19$ \\
NGC~3294     & $2.31 \pm 0.05$ & $0.77 \pm 0.11$ \\
NGC~3381     & $2.61 \pm 0.11$ & $1.01 \pm 0.19$ \\
NGC~3423     & --              & $0.55 \pm 0.13$ \\
NGC~3430     & $2.21 \pm 0.09$ & $0.66 \pm 0.08$ \\
CGCG~038-051 & --              & $1.62 \pm 0.04$ \\
IC~2828      & $> 3.09$        & --              \\
UM~439       & $< 2.09$        & $0.94 \pm 0.07$ \\
NGC~3949     & $2.18 \pm 0.04$ & $0.99 \pm 0.06$ \\
I~SZ~59      & --              & --              \\
CGCG~041-023 & --              & --              \\
SBS~1222+614 & $\sim2.92$      & --              \\
NGC~4389     & --              & $0.69 \pm 0.10$ \\
IC~3521      & $2.65 \pm 0.11$ & $0.78 \pm 0.14$ \\
UGC~9273     & --              & --              \\
MRK~475      & --              & --              \\
\hline
\end{tabular}
\label{R}
\end{table*}

\begin{landscape}
\begin{table*}
\centering
\caption{\Ha and optical morphologies of galaxies.}
\begin{tabular}{lccccc}
\hline
\hline
Name & Type & \Ha morphology & optical morphology & other features  & interaction probability \\
\hline 
UM~311        & BCD             & lopsided SF  & interacting dwarf galaxy? & lopsided radio continuum & moderate \\
IC~225        & E               & nuclear  & un-disturbed & --- & low  \\
NGC~941       & SAB(rs)c        & lopsided SF  & stellar bar, disturbed & --- &  moderate  \\
NGC~1087      & SAB(rs)c        &  lopsided SF  & misaligned \Ha and stellar bar & --- & moderate \\
UGCA~116      & Irr pec & arcs and plumes  & cometary, tidal tails & merger & high \\
UGCA~130      & Irr             & double nuclei  & cometary & past interaction & moderate \\
NGC~2799      & SB(s)m          & extra-planner \Ha regions  & lopsided, tidal tails  & close interacting pair & high \\
MRK~22        & BCD             & arcs  & double nuclei & merger  & moderate \\
IC~2524       & S               & nuclear  & --- & --- & low \\
KUG~1013+381  & BCD             & nuclear  & --- & interacting pair & moderate \\
NGC~3294      & SA(s)c          & cometary  & interacting dwarf galaxy? & --- & low \\
NGC~3381      & SB pec          & ---  & bar, lopsided & --- & low \\
NGC~3423      & SA(s)cd         & ---  & --- & disturbed radio & low  \\
NGC~3430      & SAB(rs)c        & ---  & --- & \HI tidal features & high  \\
CGCG~038-051  & dIrr            & multiple nuclei?  & asymmetric  & --- & moderate \\
IC~2828       & Im              & two nuclei?, arcs and plumes  & cometary, irregular & --- & moderate \\
UM~439        & Irr             & multiple nuclei?, arcs and plumes  & irregular &  \HI lopsided & moderate \\
NGC~3949      & SA(s)bc         & ---  & bar & offset radio emission & moderate \\
I~SZ~59       & S0              & nuclear  & elongated envelope & --- &  low \\
CGCG~041-023  & SB              & multiple nuclei?  & bar & --- & low \\
SBS~1222+614  & dIrr            & nuclear, arcs  & irregular & --- & low \\
NGC~4389      & SB(rs)bc pec    & extra-planner \Ha regions  & stellar bar, diffuse envelop & --- & moderate \\
IC~3521       & IBm             & double nuclei, arcs  & bar, diffuse envelop & --- & high \\
UGC~9273      & Im              & \Ha ring?  & lopsided & --- & moderate \\
MRK~475       & BCD             & nuclear  & lopsided & --- & low \\
\hline
\end{tabular}
\label{morphology}
\end{table*}
\end{landscape}

\addtocounter{figure}{-1}
\begin{figure*}
\centering
\epsfig{figure=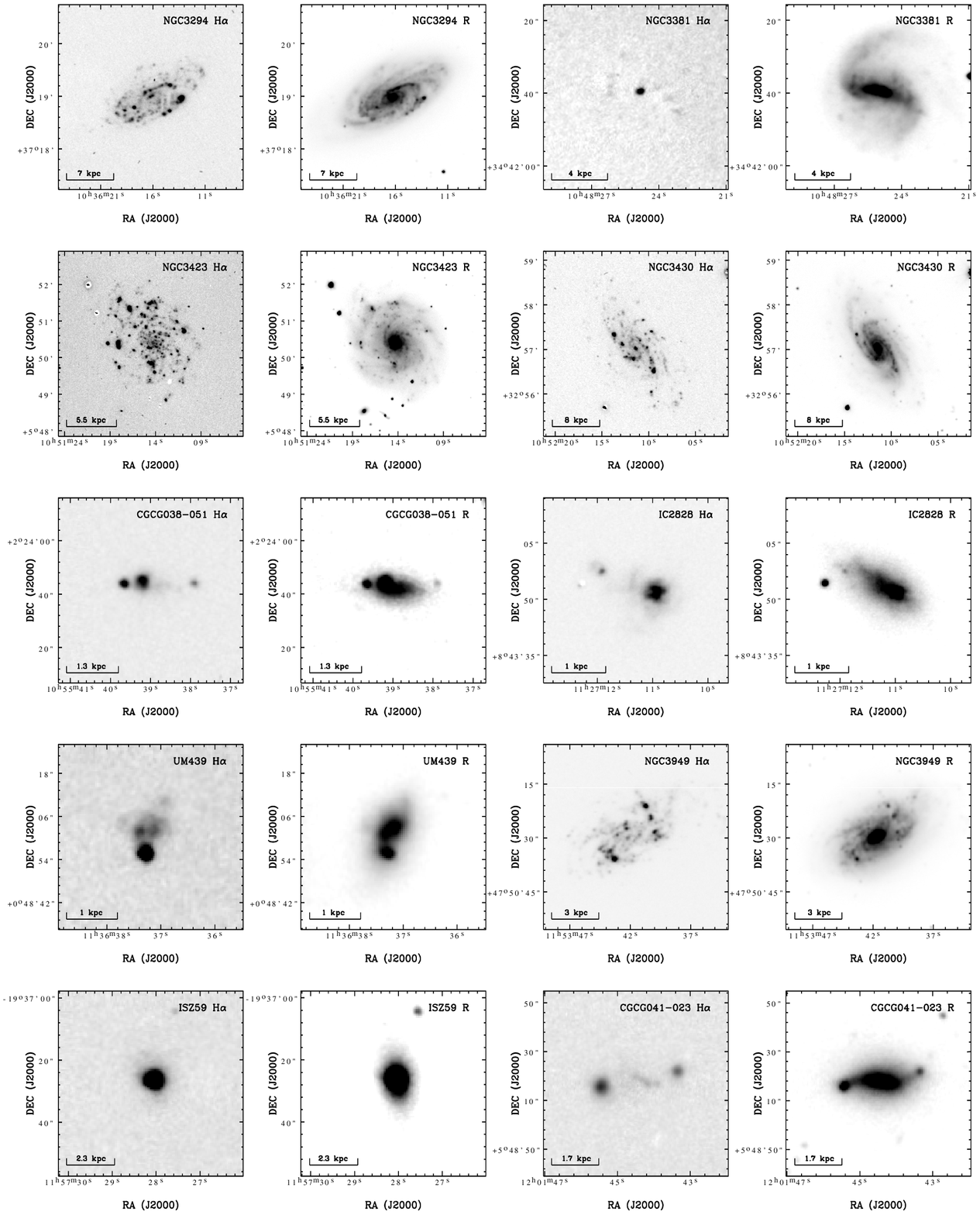,scale=0.92,angle=0}
\vspace{-1cm}
\caption{\Ha and $r$-band images of WR galaxies (continued).}
\end{figure*}
\addtocounter{figure}{-1}
\begin{figure}
\centering
\epsfig{figure=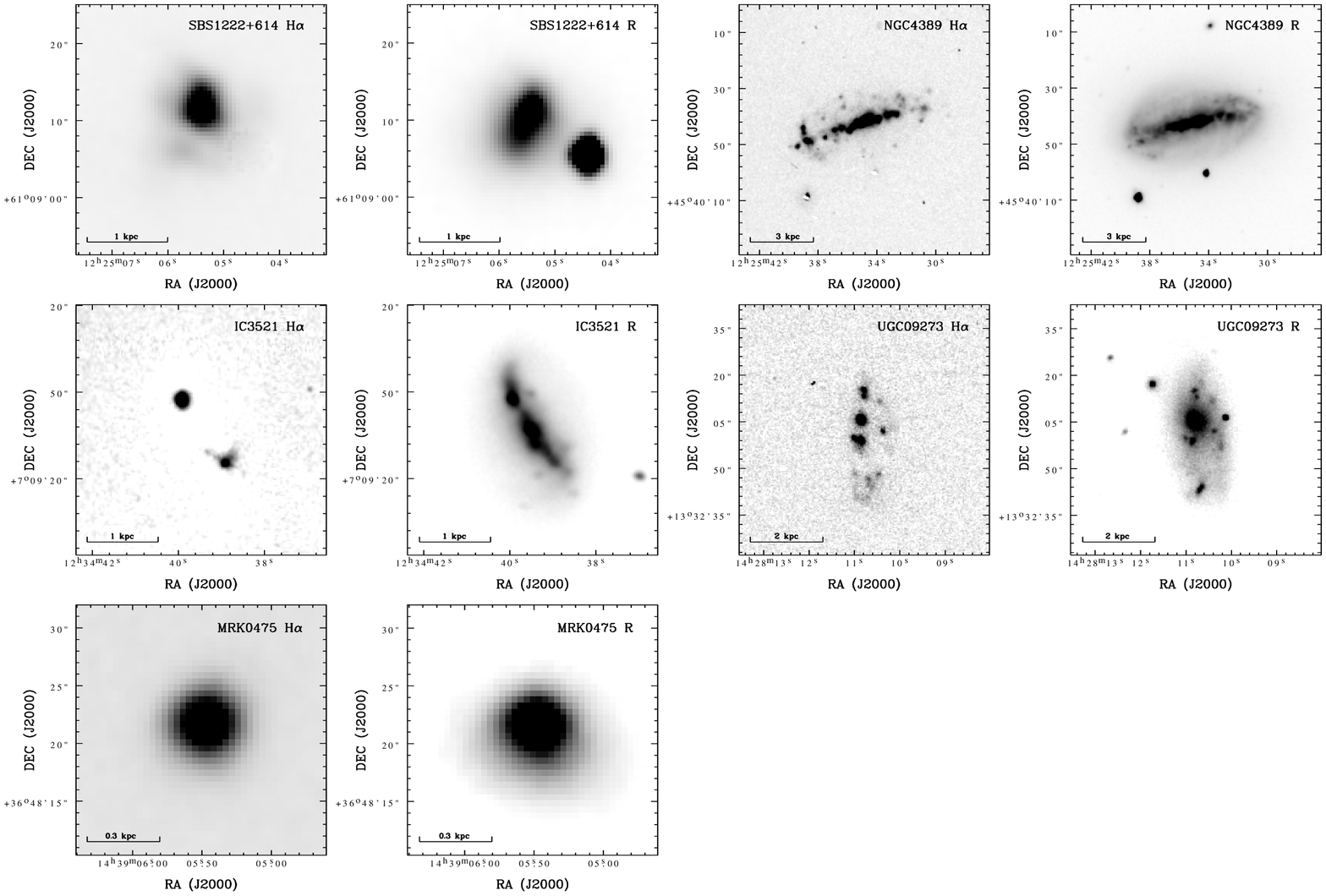,scale=0.695,angle=-90}
\vspace{-1cm}
\caption{\Ha and $r$-band images of WR galaxies (continued).}
\end{figure}

\begin{figure*}
\centering
\includegraphics[angle=270,width=0.6\linewidth]{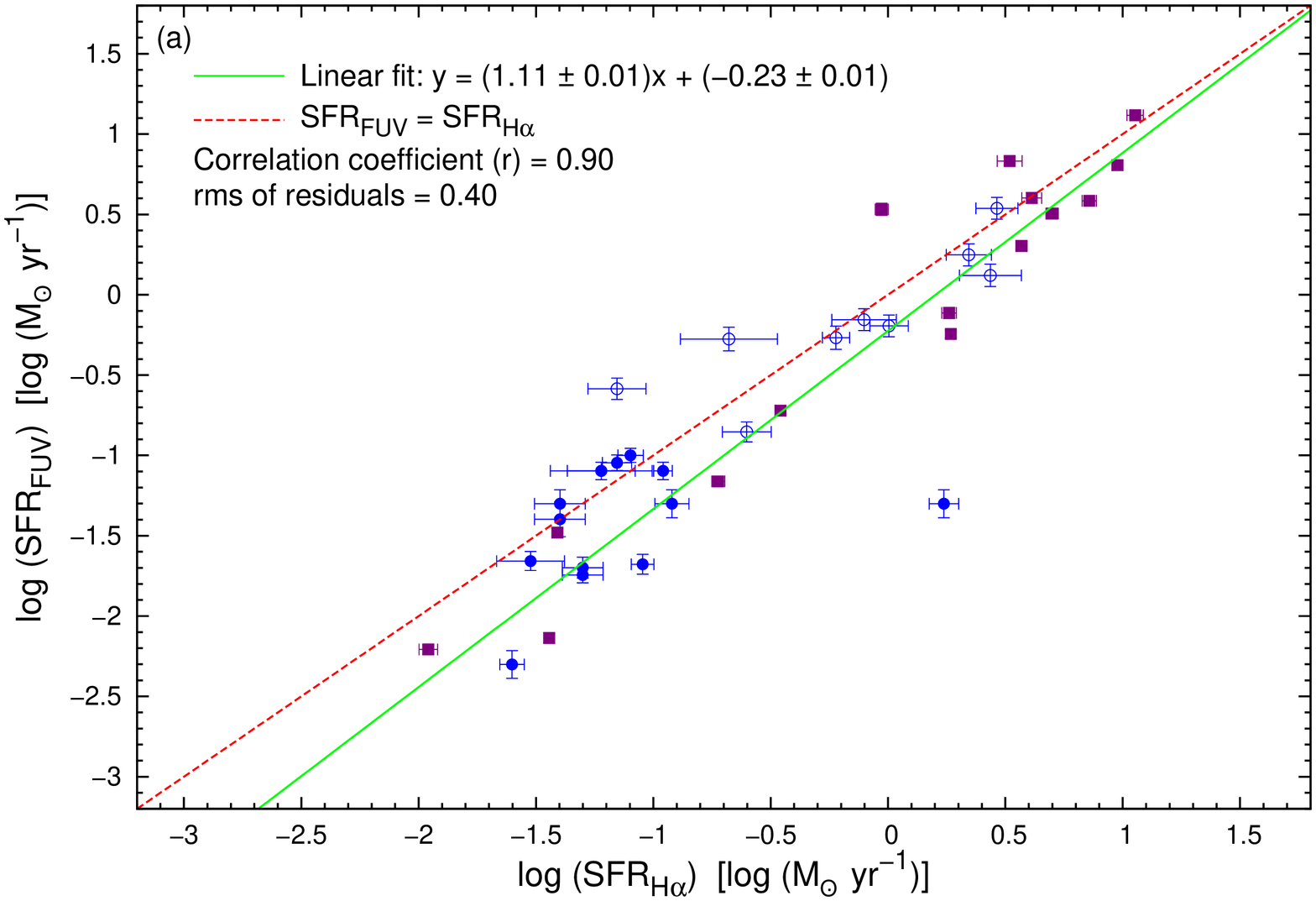}
\includegraphics[angle=270,width=0.6\linewidth]{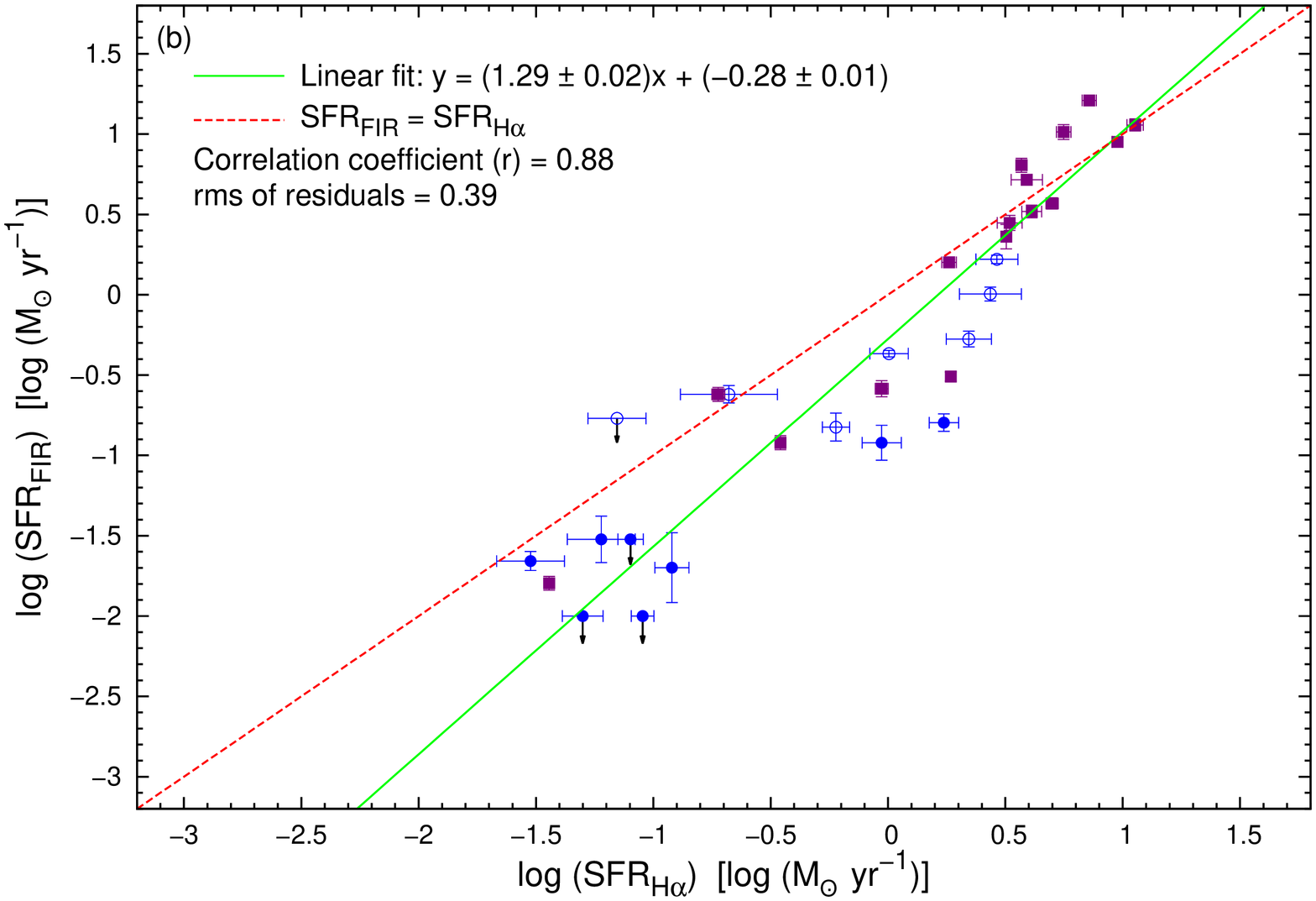}
\includegraphics[angle=270,width=0.6\linewidth]{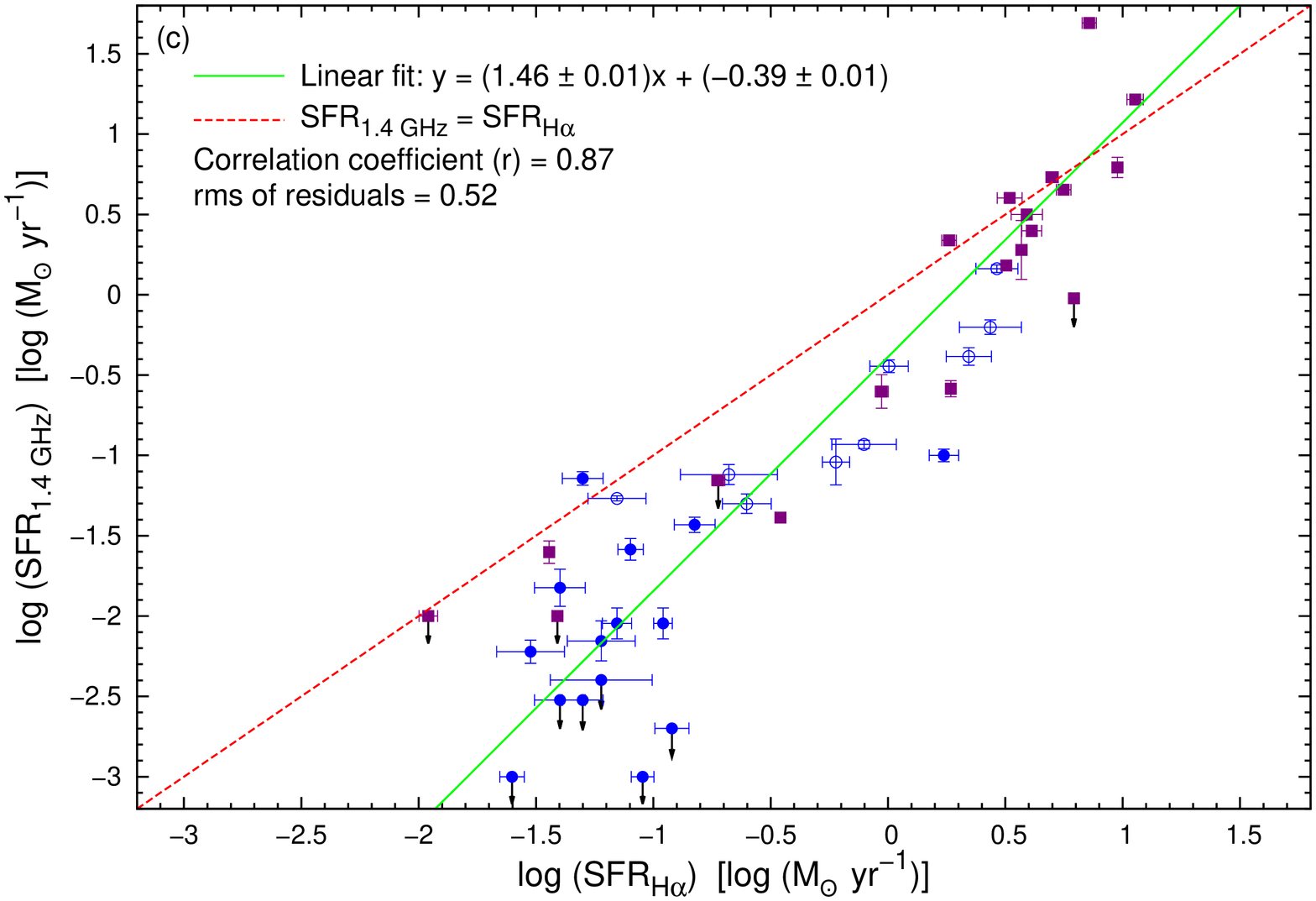}
\caption[ ]{\footnotesize{Comparison of H$\alpha$-based SFRs with the SFRs derived using luminosities in other wavebands: (a) $SFR_{FUV}$ vs. $SFR_{H\!\alpha}$ (b) $SFR_{FIR}$ vs. $SFR_{H\!\alpha}$ (c) $SFR_{1.4GHz}$ vs. $SFR_{H\!\alpha}$. The dashed lines indicate equal SFRs in two wave-bands and the solid lines are the error-weighted linear fits to the data. Small galaxies and large galaxies are designated by filled and open circles, respectively. Other data points represented by filled square are from L\'opez-S\'anchez (2010). The upper-limits to the SFR values are denoted by arrows.}}
\label{sfr1}
\end{figure*}

\begin{figure*}
\centering
\includegraphics[angle=-90,width=0.6\linewidth]{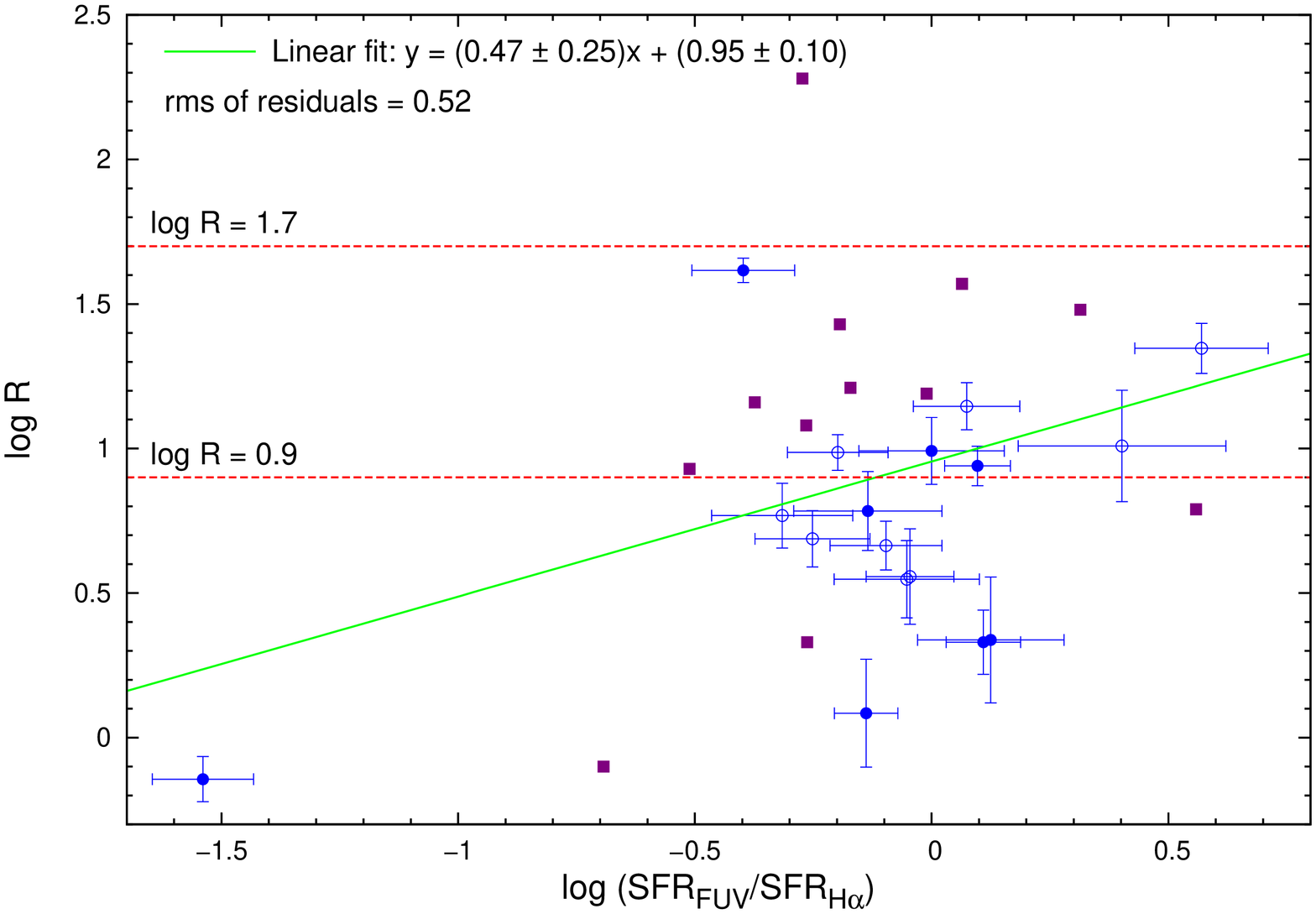}
\caption[ ]{The plot of non-thermal to thermal radio flux ratio versus FUV to \Ha SFR ratio. The symbols have the same meanings as in Figure~\ref{sfr1}. The solid line is the linear fit to the data. The dotted lines are the limits on log~R for starburst galaxies.}
\label{rplot}
\end{figure*}

\begin{figure*}
\centering
\begin{tabular}{cc}
\includegraphics[angle=270,width=0.45\linewidth]{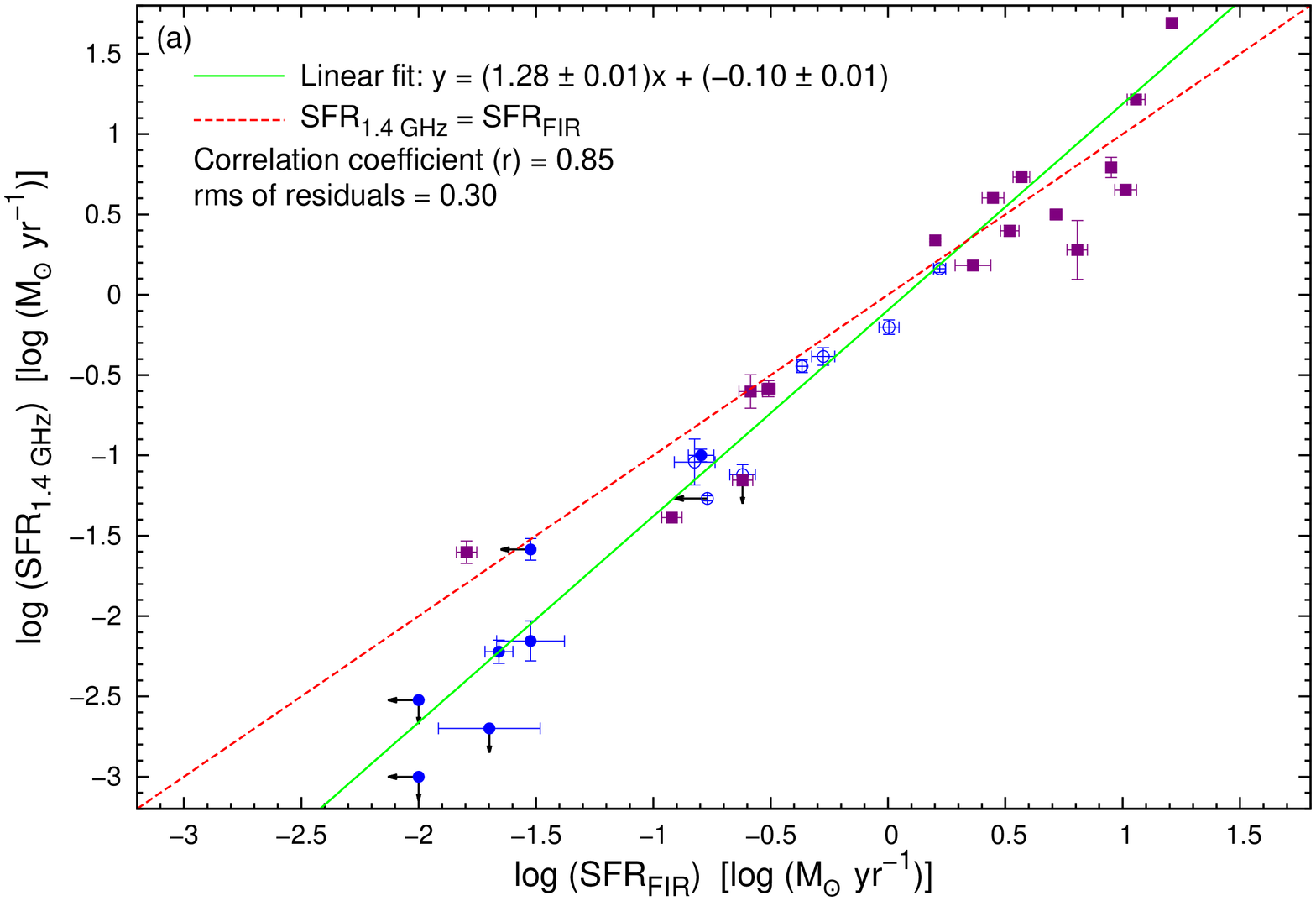} &
\includegraphics[angle=270,width=0.45\linewidth]{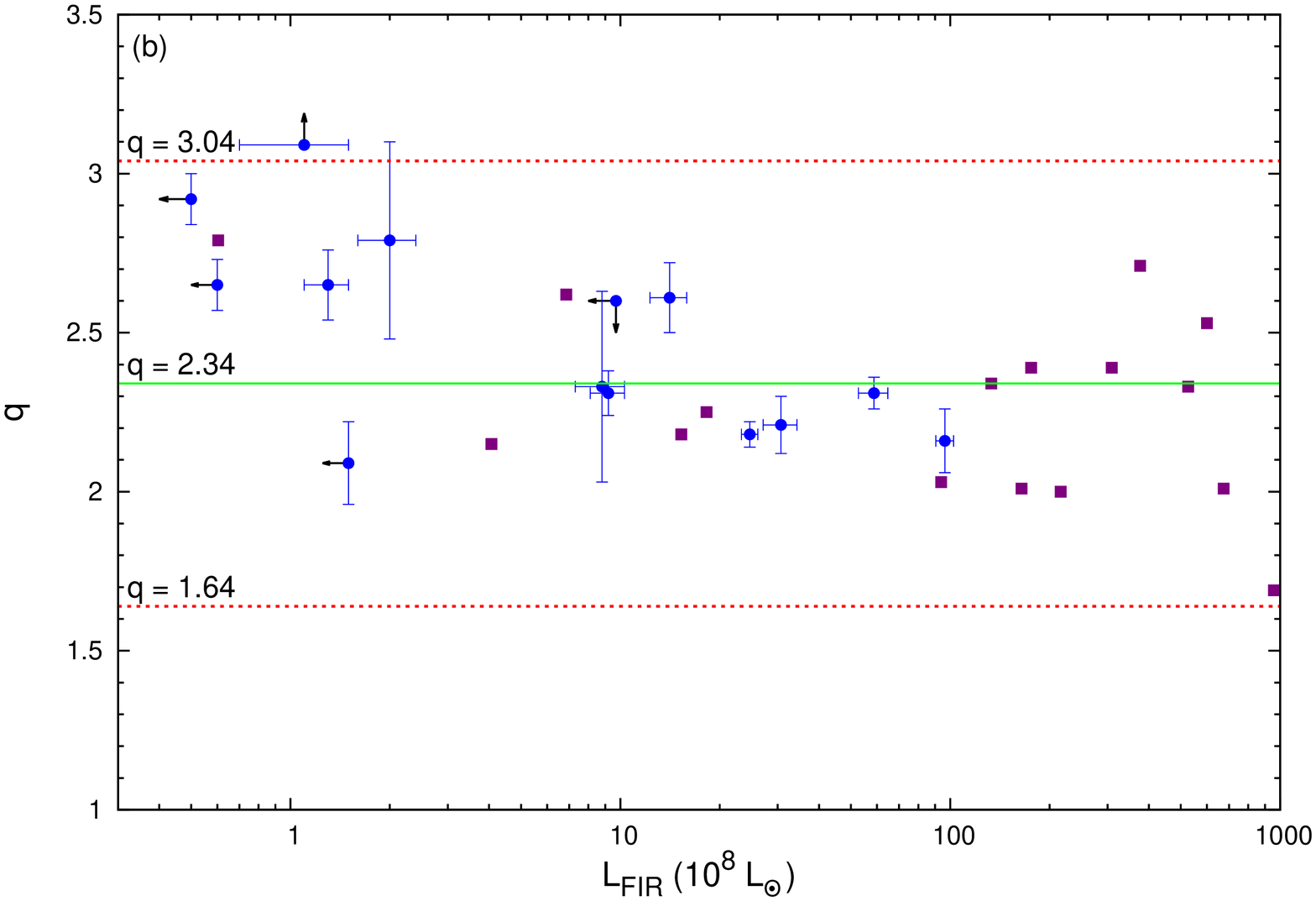} \\
\end{tabular}
\protect\caption[ ]{\footnotesize{The plots of radio based SFR versus FIR based SFR (left) and the radio-FIR correlation (right). In the left plot, symbols and lines have the same meanings as in Figure~\ref{sfr1}. In the right plot, top and bottom dotted lines are the limits of five times FIR excess and five times radio excess from the mean $q = 2.34$ (solid line) respectively. Here, symbols have the same meaning as in the left plot.}}
\label{correlation}
\end{figure*}

\end{document}